\def\be{\begin{equation}}
\def\ee{\end{equation}}
\def\ber{\begin{eqnarray}}
\def\eer{\end{eqnarray}}
\def\bers{\begin{eqnarray*}}
\def\eers{\end{eqnarray*}}

\documentclass[aps,prb,twocolumn,showpacs,groupedaddress,amsmath,amssymb]{revtex4-1}

\usepackage{dcolumn}
\usepackage{amsmath}
\usepackage{multirow}
\usepackage{setspace}
\usepackage{blindtext}
\usepackage{graphicx}
\usepackage{subfigure}
\usepackage{comment}
\usepackage{color}
\usepackage{makecell}
\usepackage{array}
\usepackage{textcomp}
     
\usepackage[T1]{fontenc}
\usepackage{hyperref}
\usepackage{braket}

\usepackage{ifthen}
\newboolean{includefigs}
\setboolean{includefigs}{true}      
\newboolean{includetext}
\setboolean{includetext}{true}     
\newcommand{\condcomment}[2]{\ifthenelse{#1}{#2}{}}

\begin{document}
\title{Thermoelectric transport and the role of different scattering processes in the half-Heusler NbFeSb}
\author{Bhawna Sahni,$^{\ast}$\textit{$^{a}$} Yao Zhao,\textit{$^{a}$} Zhen Li,\textit{$^{b}$}, Rajeev Dutt,\textit{$^{a}$}, Patrizio Graziosi,\textit{$^{c}$} and Neophytos Neophytou\textit{$^{a}$}}
\email{Bhawna.Sahni@warwick.ac.uk}

\affiliation{\textit{$^{a}$School of Engineering, University of Warwick, Coventry, CV4 7AL, UK}}
\affiliation{\textit{$^{b}$School of Materials Science and Engineering, Beihang University, Beijing 100191, China}}
\affiliation{\textit{$^{c}$Institute of Nanostructured Materials, CNR, Bologna, Italy}}

\begin{abstract}
We perform an ab initio computational investigation of the electronic and thermoelectric transport properties of one of the best performance half-Heusler (HH) alloys, NbFeSb. We use Boltzmann Transport equation while taking into account the full energy/momentum/band dependence of all relevant electronic scattering rates, i.e. with acoustic phonons, non-polar optical phonons (intra- and inter-valley), polar optical phonons (POP), and ionized impurity scattering (IIS). We use a highly efficient and accurate computational approach, where the scattering rates are derived using only a few ab initio extracted matrix elements, while we account fully for intra-/inter valley/band transitions, screening from both electrons and holes, and bipolar transport effects. \textcolor{black}{Our computed thermoelectric power-factor (PF) values show good agreement with experiments across densities and temperatures, while they indicate the upper limit of PF performance for this material. We show that the polar optical phonon and ionized impurity scattering (importantly including screening), influence significantly the transport properties, whereas the computationally expensive non-polar phonon scattering part (acoustic and non-polar optical) is somewhat weaker, especially for electrons, and at lower to intermediate temperatures.} This insight is relevant in the study of half-Heusler and other polar thermoelectric materials in general. Although we use NbFeSb as an example, the method we employ is material agnostic and can be broadly applied efficiently for electronic and thermoelectric materials in general, with more than 10x reduction in computational cost compared to fully ab initio methods, while retaining ab-initio accuracy. \\
\end{abstract}
\date{\today}

\maketitle
\section{Introduction}
In the last few decades, a new paradigm in materials science has been initiated by the widespread use of first-principles electronic structure calculations, coupled with the continuously growing power and accessibility of massively parallel supercomputers \cite{ceder1998identification,hafner2006toward,hautier2012computer,jain2016computational}. This has allowed the detailed study of material properties in a large variety of areas, as well as accurate predictions that are subsequently verified by experiment. With regards to computational studies of thermoelectric materials, and the thermoelectric power factor in particular, the electronic structure calculations are coupled to electronic transport solvers. Transport is governed by the semi-classical Boltzmann transport theory and the solution of the Boltzmann transport equation (BTE). However, this still presents significant computational challenges, making such studies impractical for realistic material systems. The difficulty lies in obtaining the charge carrier relaxation times, which are essential inputs to the BTE and are not easily obtained using current computational methods. Because of this difficulty, for electronic transport, the constant relaxation time approximation (CRTA) has typically been used to solve the BTE \cite{madsen2006boltztrap,madsen2018boltztrap2}. However, this assumption is very crude since the relaxation time typically depends on energy, momentum, band, temperature, among others, and the CRTA leads to quantitative and qualitative errors \cite{graziosi2020material,graziosi2019impact}. 
\par
One way to address this challenge is by utilizing scattering rate expressions based on deformation potentials. In 1950, Bardeen and Shockley initially suggested the deformation potential method due to acoustic phonons in the long wavelength limit, and it has since had a considerable impact on the modelling of semiconductor devices \cite{bardeen1950deformation,fischetti1996band,hong2016full}. AMSET is a recently developed program that obtains rough estimates of a ‘global’ deformation potential derived from the change in band energy due to lattice deformation (induced by phonons) \cite{ganose2021efficient}. Further advancements in density functional theory (DFT) and density-functional perturbation theory (DFPT) have enabled first-principles computations of electron–phonon interactions \cite{baroni2001phonons}, which can be used for ab initio scattering rate extraction. However, such methods are computationally very expensive. They typically require the computation and further post-processing of millions of matrix elements and need a very fine sampling of the electron and phonon momentum spaces \cite{bazhirov2010superconductivity,ponce2016epw,zhou2021perturbo}, making scaling quite challenging and impractical. In order to maintain high accuracy while reducing computational costs, various methods based on the DFPT + Wannier approach have been developed, which include averaging or integrating matrix elements across the entire Brillouin zone, such as EPA \cite{samsonidze2018accelerated} and EPICSTAR \cite{deng2020epic}. EPA takes an average over electron-phonon coupling matrix elements as well as phonon energies and thus, requires only a sparse grid. EPICSTAR also employs a sparse grid of matrix elements and adjusts them according to their phonon frequency to create a more uniform matrix element distribution. In both cases, the results are subject to the mesh discretization while convergence tests are also expensive.  
These are computationally efficient methods which provide improvement over the CRTA, although they hide detailed transport physics, especially related to intra/inter-valley/band scattering transitions. The latter is essential in band alignment strategies for TE performance improvements in multi-band/valley materials \cite{kumarasinghe2019band}. In this work, we also use an efficient formalism to compute the electronic transport properties by employing the DFPT + Wannier method and extract a limited, yet highly relevant set of matrix elements to obtain deformation potentials and determine electron-phonon scattering rates. We distinguish between all scattering processes, which allows for a complete understanding of the internal effects that determine electronic transport. We then employ our Boltzmann Transport Equation (BTE) code ElecTra \cite{graziosi2023electra}, which can account for all scattering processes individually. 
\par
We extract the TE electronic transport properties of NbFeSb, one of the more prominent TE half-Heusler materials. \textcolor{black}{We find that its transport properties are strongly influenced by electron-POP and electron-IIS scattering. At 300 K, the POP + IIS scattering mechanisms are around twice as strong in determining $\sigma$ and PF compared to ADP + ODP for p-type carriers, and five times as strong for n-type carriers, but this strength reduces with temperature. We find that the TE power factor in the p-type case peaks at 11.45 mW/mK$^2$, whereas for the n-type case at 5.92 mW/mK$^2$, which can even be achieved for an order of magnitude less density (at 6.27 x 10$^{19}/cm^3$) compared to the p-type material. Compared to experiments, we find that the calculated conductivity is around three times higher, while the Seebeck coefficient is somewhat lower, suggesting the possible presence of significant defects in experiments that lower conductivity. With regards to the PF, however, we find reasonably good agreement with existing p-type measured data, with the computational data being around two times higher.} As our approach provides valuable insights into the contribution of individual scattering mechanisms (acoustic, optical, intra- and inter-valley), it provides a way to hierarchise the scattering processes based on their strengths. Thus, we are able to provide deep understanding regarding the internal processes, and finally derive even more generalized conclusions that can be broadly applied to other half-Heusler materials and beyond. 
\par
The paper is organized as follows: In Section 2 we present the results obtained starting from the discussion of the crystal structure of NbFeSb, its electronic and phonon dispersion, the method for the extraction of deformation potentials using matrix elements obtained for p-type and n-type NbFeSb, then moving to the scattering rates and thermoelectric coefficients in section 3. In Section 4, we conclude and summarise our findings. In Section 5, we present the methods used for our transport simulation.

\section{Results}
\begin{figure*}
	\centering
	\includegraphics[scale=0.62]{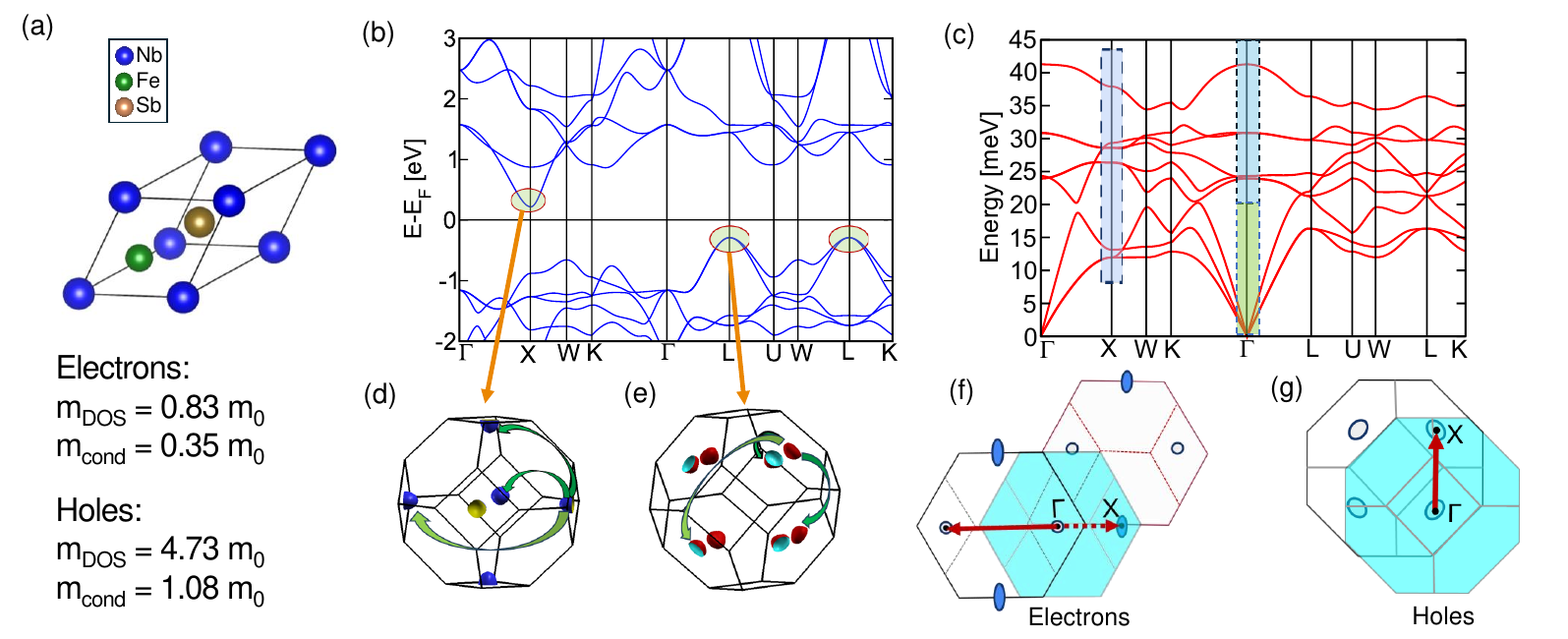}
\caption{Crystal structure: (a) Primitive unit cell for NbFeSb. (b) Electronic band structure. (c) Phonon band structure. (d),(e) Intervalley scattering within conduction band minima valleys at the X- high symmetry point and valence band maxima valleys at the L-high symmetry point respectively. (f),(g) 2D cross-section view of the Brillouin zone for electrons (transparent) and phonons (cyan) with conduction band valleys shown by blue ellipsoids and valence band valleys in grey circles, respectively, showing the path of the phonon \textbf{q}-vector along the $\Gamma$-X direction for the intervalley processes.}
	\label{structure}
\end{figure*}

Figure \ref{structure} (a) shows the cubic structure of NbFeSb, which belongs to the family of half-Heusler alloys having chemical composition XYZ, where X and Y are transition metal elements and Z is a p-block element. Half-Heusler (HH) thermoelectric (TE) materials have gathered significant research attention in the past few decades due to their thermal stability, mechanical robustness, and reasonable TE figure of merit (ZT) values. ZT is the dimensionless figure of merit that quantifies the efficiency of the energy conversion process. It is defined as $ZT$ =$(\sigma S^2)$/$(\kappa\textsubscript{e}$+$\kappa\textsubscript{L}$), where $\sigma$ is electrical conductivity, \textit{S} is Seebeck coefficient, and $\kappa\textsubscript{e} $ and $\kappa\textsubscript{L}$ are the electronic and lattice thermal conductivities, respectively. The quantity $\sigma S^2$ is called the power factor (PF). HHs are promising TE materials for use in medium to high temperature applications, which aligns well with the temperature range of many industrial waste heat sources \cite{zhu2015high,zillmann2018thermoelectric}. NbFeSb is one of the high-performance thermoelectric materials with a p-type TE figure of merit ZT > 1 \cite{he2016achieving}, which corresponds to approximately 10$\%$ of the Carnot efficiency, and it compares with the best performing materials \cite{quinn2021advances,tan2016rationally,martin2025phonon}. The investigation and clear understanding of the electronic structure and strength of the scattering mechanisms in this material, can pave the way for future optimization of TE properties in this group of materials in general. Indeed, the high TE performance of this type of materials originates mostly from their PF, which is around 3-6 mW/mK$\textsuperscript2$, and is comparable to some of the best such values across materials and operating temperatures \cite{fu2014high,yu2018unique,fu2016enhancing,ren2018ultrahigh,joshi2014nbfesb,fu2015band,kahiu2022optimized,fu2015realizing}. On the other hand, although its p-type performance is well established, very little is known about its n-type performance with only few reports available \cite{li2022synergistic,shen2019enhanced,hobbis2019structural}.

\par
\subsection{Electronic and phonon band structure}
The electronic band structure of NbFeSb is shown in Fig.\ref{structure} (b). The impact of spin-orbit coupling on the band structure of these materials has been investigated in the past and found to be negligible \cite{zhou2018large}. Thus, we proceed with the band structure without spin-orbit coupling in our calculations. Our calculations show a band gap of $E\textsubscript{g}$ = 0.51 eV which is in agreement with previous reports \cite{fu2016enhancing,kahiu2022optimized,fang2016electronic}. The atom-projected density of states of NbFeSb is shown in Fig. S1 (Supplementary Information). The conduction bands show major contribution from Nb d-orbitals only, while the valence bands show major contribution from Fe d-orbitals, followed by Nb d-orbitals and Sb p-orbitals. There is a hybridization between Nb d-orbitals and Sb p-orbitals in the energy range of 0 to -1 eV. The valence band maximum is located at the L high symmetry point and is doubly degenerate, while the conduction band minimum is at the X high symmetry point. The valence bands are flatter as compared to the conduction bands indicating higher effective masses, as noted in Fig. \ref{structure}. We have extracted the density of states effective mass ($m\textsubscript{DOS}$) and conductivity effective mass ($m\textsubscript{cond}$) for both holes and electrons using the method described in \cite{graziosi2019effective,neophytou2010large} from our EMAF code \cite{EMAFcode}. This band structure asymmetry will have some effect on the transport properties, which will be discussed in a later section.
\par
Since NbFeSb has 3 atoms in its primitive unit cell, the phonon spectrum has 9 phonon branches/modes, as shown in Fig. 1(c). The atom-projected phonon density of states is shown in Fig. S2 (Supplementary Information) which shows the major contribution of Sb atoms in low frequency regions, followed by contributions from Nb and Fe atoms. The high frequency (longitudinal optical phonon) modes show dominant contribution from Fe atoms. The highlighted green and blue regions in Fig. \ref{structure}(c) roughly show the acoustic and optical phonon mode regions considered in extracting the deformation potentials. These are regions that satisfy momentum/energy conservation to facilitate the electronic transitions. For intra-valley scattering (scattering of a carrier within its own valley), small values of wave-vector |\textbf{q}| (the zone-centre phonons) participate, which can be found in either the acoustic or optical phonon modes. For inter-valley scattering (scattering of a carrier from one valley to another), a large value of |\textbf{q}| (near zone-edge phonons) are involved due to a substantial change in momentum. The modes responsible for inter-valley scattering can also be found in both the acoustic and optical branches, but they both resemble optical modes since these modes are away from the $\Gamma$ point and are thus high in energy. Even in the case of acoustic branches participating, at those wave-vector regions they are flattened out and resemble optical modes. Schematics for the inter-valley scattering between the six conduction band minima valleys at the X point and the eight valence band maxima valleys at the L point are shown in Fig. \ref{structure}(d) and \ref{structure}(e). Figures \ref{structure}(f) and \ref{structure}(g) show the 2D cross section view for the Brillouin zone of electrons (transparent) and phonons (cyan), showing the CBM and VBM valleys with blue and grey ellipsoids respectively. The phonon wave vector \textbf{q} involved in the transition between equivalent CBM valleys and VBM valleys is along the $\Gamma$- X direction and thus the modes responsible for inter-valley scattering are shown by the highlighted region near the X-point in Fig. \ref{structure}(c). Note that this participating phonon vector direction holds for the cases of both the valence (L - L) and conduction band (X - X) transitions. In the case of the VB this is geometrically easy to identify as shown in Fig. \ref{structure}(g). For the CB, however, the vector from an X to another X-valley in the perpendicular direction is longer than the half of the Brillouin zone (the BZ for phonons), thus, we consider the equivalent final X-valley in the second electronic BZ, which happens to be in the $\Gamma$ - X direction as shown in Fig. \ref{structure}(f).     

\subsection{Deformation potential extraction method}
For acoustic phonons (in the long wavelength limit, i.e. small wave-vector |\textbf{q}|), the atomic displacements can cause the crystal to deform. Such deformations alter the electronic energies at various locations inside the Brillouin zone. Deformation potentials are the parameters that characterise these changes in electronic energies, brought on by static distortions of the lattice \cite{bardeen1950deformation}. The deformation potentials can be more accurately determined by calculating the electron-phonon coupling matrix elements, from which we can determine the electron-phonon coupling strength, and using Fermi’s Golden rule determine the scattering rates. The scattering matrix elements are the probability amplitude for scattering from an initial state $\ket{\psi_{n\textbf{k}}}$ to a final state $\ket{\psi_{m\textbf{k}+\textbf{q}}}$ due to a perturbing potential  $\delta_{\textbf{q}\nu}V$ associated with a phonon on branch $\nu$, frequency $\omega_{\textbf{q}\nu}$ and wave vector \textbf{q}, are defined as \cite{savrasov1994linear,liu1996linear,Allen1984}: 
\begin{equation}
    M_{mn\nu} (\textbf{k},\textbf{q})=\bra{\psi_{n\textbf{k}}}\delta_{\textbf{q}\nu}V\ket{\psi_{m\textbf{k}+\textbf{q}}}
    \label{eqn1}
\end{equation}
The acoustic deformation potential (ADP) can be calculated by taking the slope of the matrix elements with respect to the phonon wave vector,  $D\textsubscript{ADP}$ = ($M_{mn\nu} (\textbf{k},\textbf{q})$ )/(|\textbf{q}|) \cite{lundstrom2002fundamentals}. The optical phonons on the other hand, give rise to microscopic distortions which can be regarded as an internal strain within the unit cell. The optical deformation potential (ODP) is the value of the matrix element itself, $D\textsubscript{ODP}$ = ($M_{mn\nu} (\textbf{k},\textbf{q})$ ). The acoustic and optical deformation potential scattering mechanisms are of short-range in nature, and they do not depend explicitly on the phonon wave-vector \textbf{q}.
\par
In the case of polar crystals, the long wavelength (small |\textbf{q}|) longitudinal optical phonon modes (LO) induce an oscillating polarization, generating a macroscopic electric field \cite{born1996dynamical}. This electric field couples with the carriers and is known as the Frohlich interaction \cite{frohlich1954electrons}, which is long-range in nature. The \text{Fr\"ohlich}
 interaction becomes stronger for smaller |\textbf{q}| and it diverges as |\textbf{q}| becomes zero. The long-range component of the electron-phonon coupling matrix elements due to the \text{Fr\"ohlich}
 interaction is defined as \cite{frohlich1954electrons}:
\begin{align}
M_{mn\nu}^L(\textbf{k}, \textbf{q}) &= i \frac{m_0^{1/2} e^2}{\Omega \epsilon_0} 
\sum_k M_k^{-1/2} \notag \\
&\quad \times \sum_{\textbf{G} \neq -\textbf{q}} 
\frac{(\textbf{q} + \textbf{G}) \cdot \mathbf{Z}_k^* \cdot \mathbf{e}_{k\nu}(\textbf{q})}{(\textbf{q} + \textbf{G}) \cdot k_\infty \cdot (\textbf{q} + \textbf{G})} \notag \\
&\quad \times \braket{u_{m\textbf{k}+\textbf{q}} | e^{i(\textbf{q} + \textbf{G})\cdot \mathbf{r}} | u_{n\textbf{q}}}
\label{eqn2}
\end{align}
where $\Omega$ is the unit cell volume, $m\textsubscript{0}$ is the sum of the masses of all the atoms in the unit cell, M$_k$ is the mass of atom k, $\epsilon_0$ is the vacuum permittivity, \textbf{G} indicates the reciprocal lattice vector, $e_{\textbf{k}\nu}(\textbf{q})$ indicates a normalized vibrational eigenmode within the unit cell, k$_\infty$ is the high-frequency dielectric constant and \textbf{Z}$_k^*$ is the Born effective charge tensor. A polar phonon mode can include both, a short-range (deformation potential) part, and a long-range (POP) part \cite{park2020long}. The long-range component needs to be separated from the short-range component to deal with the polar singularity as the wave vector approaches zero. For calculating acoustic and optical deformation potentials for polar materials, the long-range part of the matrix elements is subtracted from the total matrix elements to get the short-range part of the matrix elements \cite{verdi2015frohlich}. The short-range component of matrix elements is then used to calculate $D\textsubscript{ADP}$ and $D\textsubscript{ODP}$. 
\par
The overall $D\textsubscript{ADP}$ value is calculated after averaging the longitudinal and transverse acoustic deformation potentials as:  
\begin{equation}
    D\textsubscript{ADP}= v_s \sqrt{\frac{D\textsubscript{ADP,LA}^2}{v_l^2 }+\frac{D\textsubscript{ADP,TA}^2}{v_t^2}},
\label{eqn3}
\end{equation}
where $v_s$, $v_l$ and $v_t$ are the average, longitudinal, and transverse sound velocities, respectively. \textcolor{black}{Note that ElecTra can take into account the contributions of each acoustic mode separately as well. However, we find that considering the averaged deformation potential in this way provides almost identical results for the acoustic-limited power factor calcaultions, thus we proceed with the averaged value to keep the number of input parameters lower (see Fig. S3 in the SI for details).}The $D\textsubscript{ODP}$ value is calculated by taking the average of the short-range component of all optical modes for small \textbf{q}-vector values near the $\Gamma$ point as:
\begin{equation}
D{\textsubscript{ODP}} = \sqrt{\omega{\textsubscript{ODP}} \sum_{\nu} \frac{|M_{mn\nu}(\textbf{k}, \textbf{q}) - M_{mn\nu}^L(\textbf{k}, \textbf{q})|^2}{\omega_{\nu}}}
\label{eqn4}
\end{equation}
where, $\omega\textsubscript{ODP}$ is the average value of optical phonon energy. \textcolor{black}{Both of these deformation potentials are calculated along different crystallographic directions to take anisotropy into account, and then averaged to calculate the overall deformation potentials.} 

We follow this process of subtracting the long-range part of the matrix elements from the total matrix element to obtain the short-range part for all modes for generality in the computational approach. Of course, we know that only two of the modes are polar and include long-range parts \cite{li2024efficient}. In the other modes, the long-range part is zero, but it is more convenient for the numerical automation of the process, since we do not identify which ones are polar to begin with. The POP scattering rate itself, is then computed using the Frohlich formula for POP scattering including screening, as described in the methods section, whereas the average mode frequency is used for the interaction.
\par
For inter-valley scattering, as discussed earlier, phonons with large wave-vectors |\textbf{q}|, scatter electrons from one valley to another valley of the same or different band. The zone-edge phonons away from the $\Gamma$ point are the ones which mediate these scattering events. The inter-valley deformation potential is obtained again by considering the contribution from all the modes and taking an average as \cite{li2024efficient}: 
\begin{equation}
    D{\textsubscript{IVS}} = \sqrt{\omega{\textsubscript{IVS}} \sum_{\nu} \frac{|M_{mn\nu}(\textbf{k}, \textbf{q}) - M_{mn\nu}^L(\textbf{k}, \textbf{q})|^2}{\omega_{\nu}}}
    \label{eqn6}
\end{equation}
where $\omega\textsubscript{IVS}$ is the average value of frequency for all phonons participating in the inter-valley scattering process.

\subsection{Matrix elements for NbFeSb} 
The first step for the calculation of the deformation potentials is identifying the band extrema and all different possible transitions (intra/inter-band/valley) from the electronic band structure of the material. Using the wavevectors that facilitate those transitions, the participating phonon modes that satisfy energy/momentum conservation are identified and then the electron-phonon coupling matrix elements for all of them are computed. In this section, we first compute the deformation potentials for transitions in the valence band (VB) of NbFeSb and afterwards in the conduction band (CB).
\begin{figure*}
	\centering
	\includegraphics[scale=0.36]{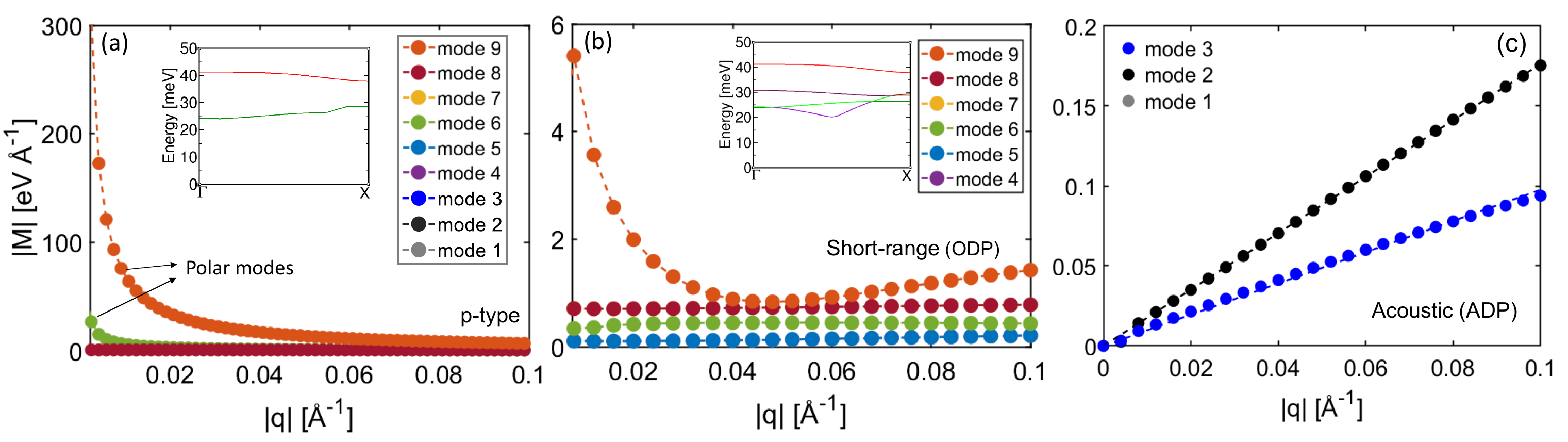}
\caption{Scattering matrix elements for intravalley deformation potential extraction for holes: (a) For all phonons (two modes show polar behaviour) along the $\Gamma$ - X direction. (b) Short range part of the matrix elements for optical and (c) acoustic phonons for scattering from the valence band maxima (VBM to VBM). The insets in (a) and (b) show the frequencies corresponding to the polar optical and non-polar optical modes, respectively.}
	\label{ML_ptype}
\end{figure*}

\subsubsection{Valence band (p-type) NbFeSb }
We start with the VB of NbFeSb. First, we seek to extract scattering information for intra-valley processes, that involve small |\textbf{q}| phonons: i) by polar optical phonons, ii) by non-polar optical phonons, and iii) by acoustic phonons. Since the VBM has two degenerate bands (labelled here B1, B2) at the L-point, we calculate the matrix elements for scattering associated with all four possible transition combinations, i.e. from B1$\rightarrow$B1, B1$\rightarrow$B2, B2$\rightarrow$B1, B2$\rightarrow$B2. After subtracting the long-range part, we first calculate the ADP and ODP values using Eqns. 3 and 4, respectively, along different crystallographic directions for the four scattering processes mentioned above. Interestingly, the matrix elements and deformation potential values are the same for transitions with B1 as the final state i.e. values for B2$\rightarrow$B1 and B1$\rightarrow$B1 are identical. In the same way, transitions with the B2 as the final state i.e. B1$\rightarrow$B2 and B2$\rightarrow$B2 are also identical. This is the same as observed for transitions in the valence band of Si as well \cite{li2021deformation}.  We then average the ADP and ODP values from these four scattering processes along different directions and calculate the overall ADP and ODP value using Eqn. 3 and 4 respectively. 
\par
Figure \ref{ML_ptype}(a) shows the scattering matrix elements for intra-valley transitions (small \textbf{q}-vector) for B1$\rightarrow$B1 along the $\Gamma$ - X direction (see supplementary Fig. S5 for the matrix elements in the other directions, which are of similar trend and amplitude). There are two polar modes due to the two longitudinal optical phonon modes (mode 6 and 9) as shown in Fig. \ref{ML_ptype}(a), recognised by their 1/|\textbf{q}| trend. The rest of the matrix elements are much lower in amplitude. The inset of Fig. \ref{ML_ptype}(a) shows the frequencies of these two polar modes along the same direction. The average value of the polar optical phonon energy is calculated along various high-symmetry directions. Then those values from all polar modes are averaged to extract a single dominant frequency for the overall polar optical phonon scattering process. This is determined to be $\hbar\omega$ = 32 meV here, and that is used in the equation for the Frohlich interaction (see methods).  
The short-range part of these matrix elements for the six optical modes is shown in Fig. \ref{ML_ptype}(b) after the long-range part is subtracted out. The average optical phonon energy is determined to be $\hbar\omega$ = 29 meV. 
\par
The acoustic modes in Fig. \ref{ML_ptype}(c) shows the short-range part of the matrix elements corresponding to the longitudinal and transverse acoustic modes. The intra-valley deformation potentials for all four scattering processes (between the two bands) along different high-symmetry directions are listed in Table S1 (Supplementary Information). The calculated ADP and ODP values along different high-symmetry directions are shown in Table 1. \textcolor{black}{The overall ODP and ADP values for holes turn out to be $D\textsubscript{ODP}$ = 2.8 eV$\AA^{-1}$ (with an overall phonon energy of $\hbar\omega$ = 29 meV) and $D\textsubscript{ADP}$ = 2.5 eV, respectively.} \textcolor{black}{The overall deformation potentials are calculated as:}
\begin{equation}
D = \sqrt{\frac{n_{\Gamma-L} D_{\Gamma-L}^2 + n_{\Gamma-K} D_{\Gamma-K}^2  + n_{\Gamma-X} D_{\Gamma-X}^2 }{n_{\Gamma-L} + n_{\Gamma-K} + n_{\Gamma-X}}}
\label{eqn5}
\end{equation}
\textcolor{black}{The number of the equivalent crystallographic orientations for a face-centered cubic (FCC) lattice are $n_{\Gamma-L} = 8$, $n_{\Gamma-X} = 6$, and $n_{\Gamma-K} = 12$. Note that we used sampling along the <111>, <100> and <110> crystal directions. This covers the majority of the volume of the BZ (ending up on the hexagonal and square surfaces on the BZ, and the edges around those surfaces, respectively - see Fig. S4 in the SI). We have also computed the deformation potentials along the path between $\Gamma$ and $W$, which is also towards the edge of the BZ in between the hexagonal and square surfaces. This gives similar values as the $\Gamma$ and $K$ direction (which is towards the edges as well). We exclude this direction from the sampling, as is covered by the $\Gamma$ and $K$ and as it is not one of crystal directions.}

\begin{table}[h]
\centering
\begin{tabular}{|c|c|c|c|}
    \hline
    \multicolumn{4}{|c|}{Holes} \\  
    \hline
    & $\Gamma$-L  & $\Gamma$-X & $\Gamma$-K  \\
    $D_{ADP}$ (eV) & 2.38 & 2.16 & 2.76 \\ 
    $D_{ODP}$ (eV\AA$^{-1}$) & 2.58 & 2.13 & 3.11 \\ 
    \hline
\end{tabular}
\caption{The acoustic and optical deformation potential values for holes along different high symmetry directions.}
\label{D_holes}
\end{table}

\begin{figure}[h!]
	\centering
	\includegraphics[scale = 0.36]{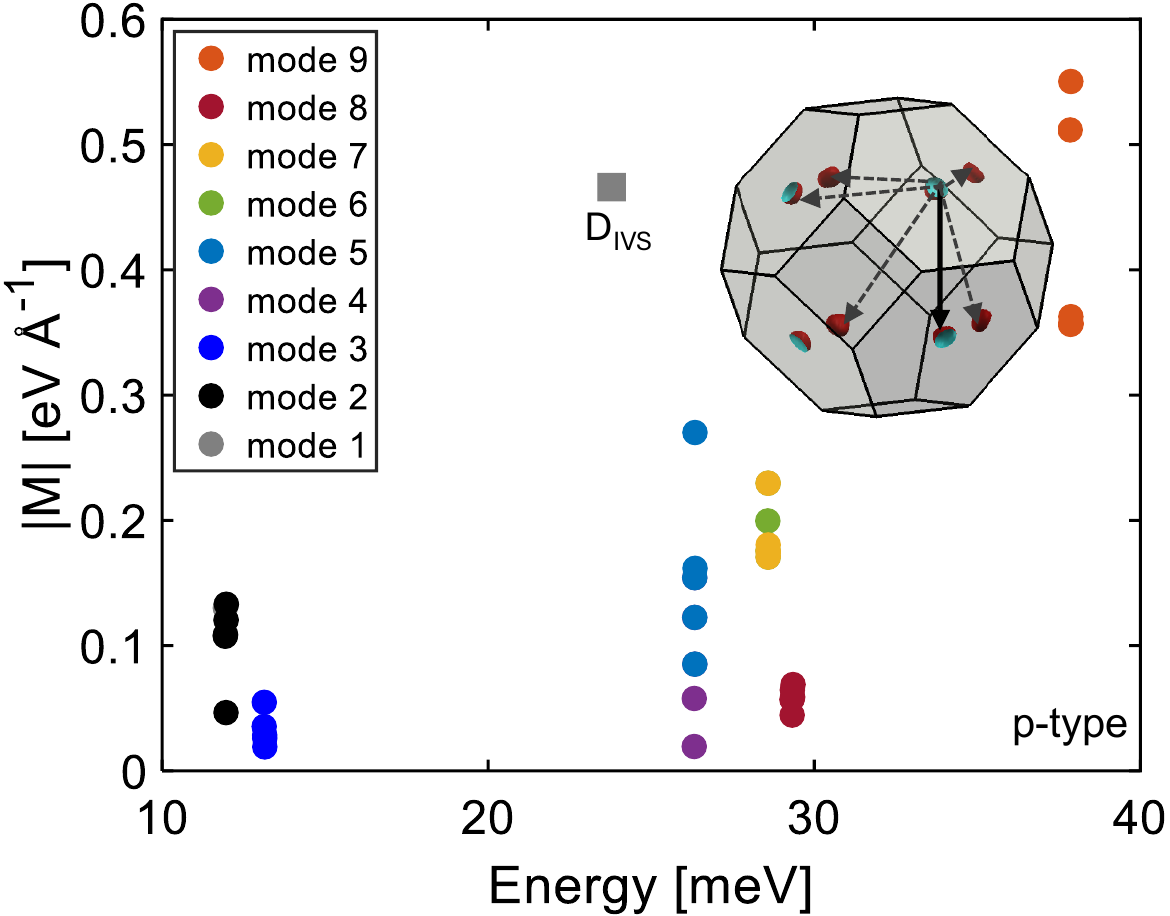}
\caption{Scattering matrix elements for intervalley deformation potential extraction for holes. The data correspond to transitions from an initial VBM valley (on the L-point) to any other equivalent VBM valley (on the L-point), as shown by solid black arrow in BZ. The other equivalent transitions are shown by the dashed arrows. The contributions of the different modes are shown by different colors. The brown square shows the overall value of intervalley deformation potential for these transitions.}
	\label{IVS_ptype}
\end{figure}
\begin{figure*}[htbp!]
	\centering
	\includegraphics[scale=0.36]{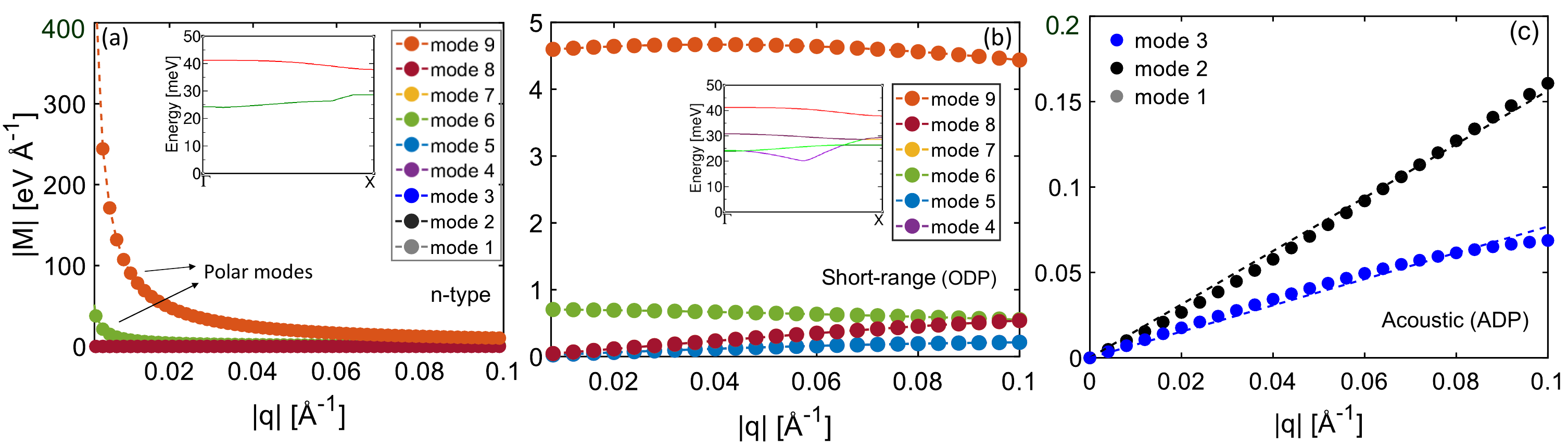}
\caption{Scattering matrix elements for intravalley deformation potential exaction for electrons: (a) For all phonons (two modes show polar behaviour) along the $\Gamma$-X direction. (b) Short range part of matrix elements for optical and (c) acoustic phonons for scattering from the conduction band minima (CBM to CBM). The insets in (a) and (b) show the frequencies corresponding to the polar optical and non-polar optical modes, respectively.}
	\label{ML_ntype}
\end{figure*}

Next, we consider the inter-valley scattering processes. Since the VBM resides on the L high-symmetry point, the Fermi surface of holes contains eight half L-valleys in the first Brillouin zone (BZ). Due to the cubic crystal symmetry, only one unique type of inter-valley transition from any given initial VBM valley to other equivalent VBM valley can be identified, as shown by the solid black arrow in the BZ of Fig. \ref{IVS_ptype}. The other equivalent transitions are shown by the dashed grey color arrows. Figure \ref{IVS_ptype} shows the matrix elements from all phonon modes for that transition, identified by different colors for different phonon modes. We considered a few points in the vicinity of the band extrema ($\sim$ 5 points on each final valley – from which we take the average value) and thus, there are 45 points from 9 phonon modes in Fig. \ref{IVS_ptype}, corresponding to transition from one valley to another. The brown square in Fig. \ref{IVS_ptype} shows the overall inter-valley matrix element (and deformation potential) value for these transitions \textcolor{black}{for VBM (B1)}. Note that the energies at which the different points appear, correlated very well from the energy of the phonon bands under the highlighted regions in the phonon spectrum of Fig. \ref{structure}. The $D\textsubscript{IVS}$ value (\textcolor{black}{average of B1$\rightarrow$B1 (0.48), B1$\rightarrow$B2 (0.56)}) is 0.52 eV\AA$^{-1}$ as computed using Eqn. 6 including all nine modes (one averaged value for each mode) with the average frequency value $\hbar\omega$ $\sim$ 24 meV.

\subsubsection{Conduction band (n-type) NbFeSb}
For the CB of NbFeSb, we proceed in the same manner as for the VB above. Unlike the VB, the CB has only one band at the X-point, thus we extract the matrix elements for scattering from that band to itself alone.  Figure \ref{ML_ntype}(a) shows those scattering matrix elements for intra-valley transitions (small \textbf{q}-vector) along the $\Gamma$-X direction (see supplementary Fig. S5 for the matrix elements in other directions, which are of similar trend and amplitude). The two polar optical modes are also evident in Fig. \ref{ML_ntype}(a), with their average phonon energy (single dominant frequency for the overall polar optical phonon scattering process) being $\hbar\omega$ = 32 meV, as also computed earlier for the CB. The short-range part of these matrix elements for the six optical modes is shown in Fig. \ref{ML_ntype}(b). The average optical phonon energy is determined to be $\hbar\omega$ = 29 meV (same as above for the CB). Figure \ref{ML_ntype}(c) shows the short-range part of the matrix elements corresponding to the longitudinal and transverse acoustic modes. The calculated intra-valley ADP and ODP values along different high-symmetry directions for electrons are shown in Table 2. \textcolor{black}{ The overall ODP and ADP values calculated turn out to be $D\textsubscript{ODP}$ = 3.3 eV\AA$^{-1}$ (with overall phonon energy of  $\hbar\omega$ = 29 meV) and $D\textsubscript{ADP}$ = 4.7 eV respectively.}
\begin{table}[h]
\centering
\begin{tabular}{|c|c|c|c|}
    \hline
    \multicolumn{4}{|c|}{Electrons}  \\
    \hline
    & $\Gamma$-L  & $\Gamma$-X & $\Gamma$-K  \\
    $D_{ADP}$ (eV) & 1.93 & 2.56 & 6.51 \\
    $D_{ODP}$ (eV$\AA^{-1}$) & 3.7 & 3.88 & 2.51  \\
    \hline
\end{tabular}
\caption{The acoustic and optical deformation potential values for electrons along different high symmetry directions.}
\label{D_electrons}
\end{table}

\begin{figure}[h!]
	\centering
	\includegraphics[scale = 0.4]{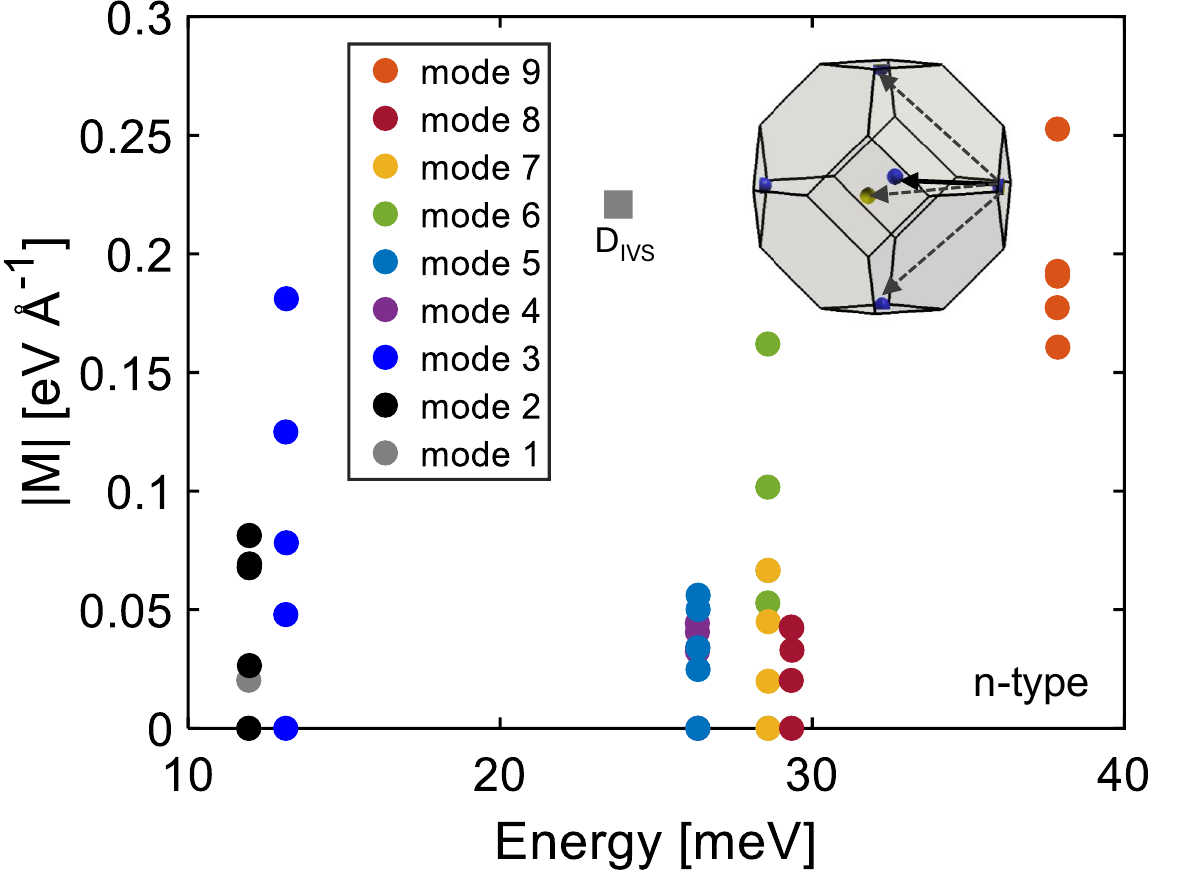}
\caption{Scattering matrix elements for intervalley deformation potential for electrons: The data show transitions from an initial CBM valley (on the X-point) to any other equivalent CBM valley (on the X-point) by the solid black arrow in the BZ. The other equivalent transitions are shown by the dashed arrows. The contributions of the different modes are shown by different colors. The brown square shows the overall value of intervalley deformation potential for these transitions.}
	\label{IVS_ntype}
\end{figure}
\par
For inter-valley scattering, the CBM is at the X high-symmetry point, so the Fermi surface of electrons contains six half X-valleys in the first Brillouin zone (BZ). One unique type of inter-valley transitions from any given initial CBM valley to other equivalent CBM valleys can be identified, as shown by the solid black arrow in the BZ of Fig. \ref{IVS_ntype}. The other equivalent transitions are shown by the dashed grey color arrows. The matrix elements for the transitions facilitated by all phonon modes are shown in Fig. \ref{IVS_ntype}, where each color corresponds to a different mode. Again, we consider 5 points in the vicinity of each of the band extrema, resulting in 45 points in Fig. \ref{IVS_ntype} for transitions from one valley to another. The brown square shows the overall inter-valley deformation potential value for this transition. The $D\textsubscript{IVS}$ values are computed to be $D\textsubscript{IVS}$ = 0.22 \AA$^{-1}$  with the average frequency value of $\hbar\omega$ $\sim$ 24 meV (as computed above for the VB as well).

\section{Scattering rates and thermoelectric transport}
\begin{figure*}[!htbp]
    \centering
    \begin{minipage}{0.7\textwidth}
        \includegraphics[width=\textwidth]{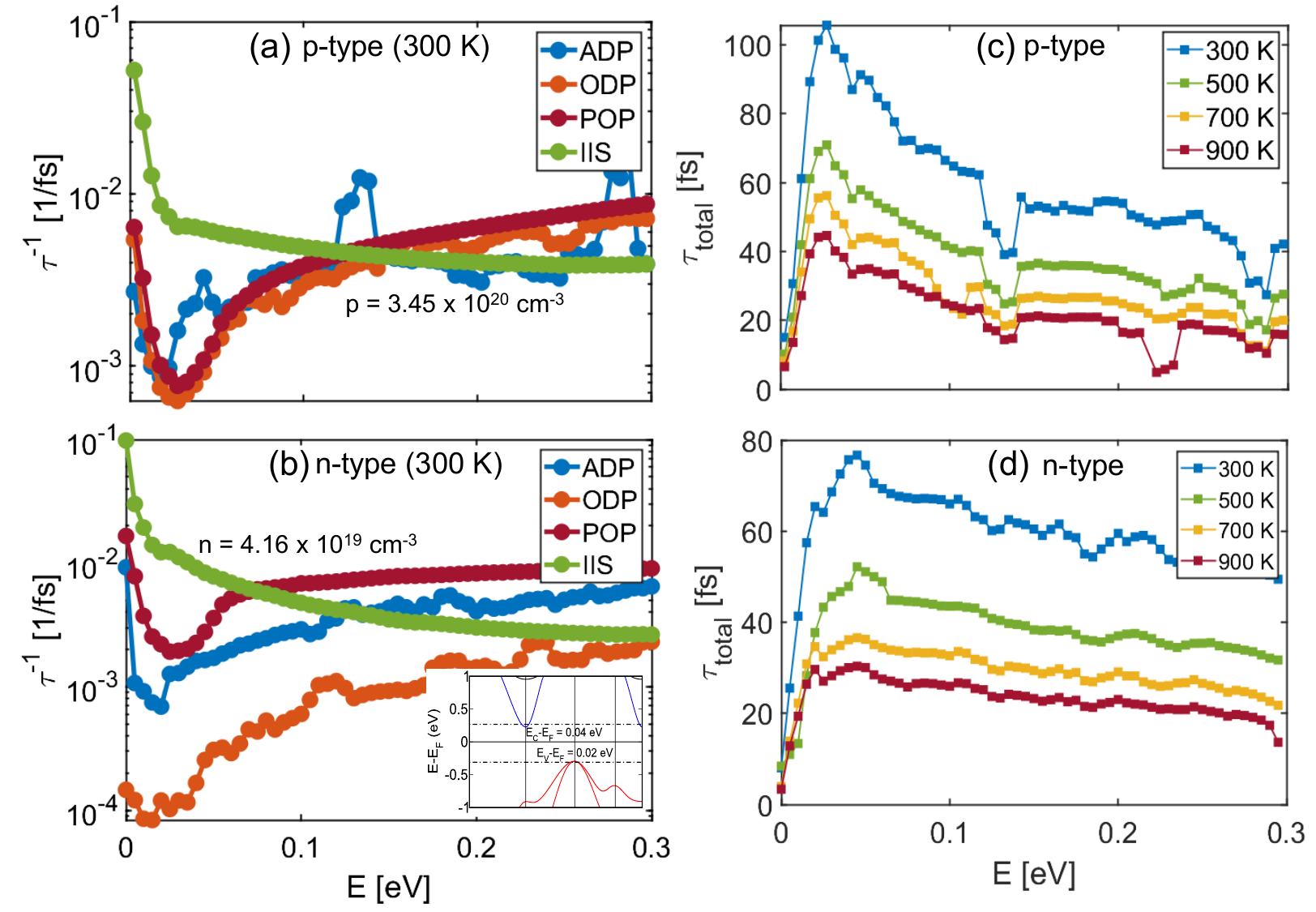}
    \end{minipage}%
    \hfill
    \begin{minipage}{0.25\textwidth}
        \vspace*{\fill} 
        \caption{Scattering rates for NbFeSb: (a) Scattering rates related to different mechanisms (as denoted) for p-type and (b) n-type NbFeSb at 300 K. The inset in (b) shows the Fermi level at which POP scattering rates are shown (i.e. where maximum power-factor is obtained). (c) and (d) show the total carrier relaxation time as a function of energy and temperature for holes and electrons, respectively.}\label{rates_tau}
        \vspace*{\fill}
    \end{minipage}
\end{figure*}
The transport calculations are performed using our open-source Boltzmann Transport Equation solver (BTE) code ElecTra \cite{graziosi2023electra}. The expressions for individual scattering rates are described in the computational methods section below. The BTE simulator takes the deformation potentials and the dominant frequency for polar optical phonons as inputs for the electronic transport calculations. Table S2 (in the Supplementary Information file) lists the other required input parameters, such as dielectric constants and mass densities, which are also obtained through first-principles calculations. In the calculation we do not separate p-type and n-type transport, but treat the entire band structure as one entity, including bipolar effects, and full treatment of screening, i.e. both electrons and holes contribute to screening, irrespective of the Fermi level position \cite{graziosi2020ultra}. Thus, the calculations are performed under the assumption of a complete bipolar transport, integrating the complete transport distribution function TDF (see methods) across the entire energy range. Note that although it is computationally more convenient and easier to perform unipolar transport calculations and afterwards combine the two parts, in the presence of IIS this is not possible, and the full bipolar computation needs to be performed in order to capture the impurity density and screening correctly \cite{graziosi2020ultra}.
\par
 Figures \ref{rates_tau}(a) and \ref{rates_tau}(b) show the individual scattering rates for p-type and n-type NbFeSb at 300 K when the Fermi level is placed in the vicinity of the respective band edge, as shown in the inset of Fig. \ref{rates_tau}(a); this is the region where the PF peaks, as we show later on. \textcolor{black}{For p-type, the ADP, ODP and POP scattering rates are comparable, while IIS is the dominant scattering mechanism. For n-type, ADP and ODP scattering rates are lower while POP and IIS are the dominant scattering mechanisms.} The total relaxation time values obtained by Matthiessen’s rule (explained in the methods section), for both holes and electrons for various temperatures, are shown in Fig. \ref{rates_tau}(c) and \ref{rates_tau}(d) respectively, again when the Fermi level is placed at the vicinity of the band edges. The relaxation time values vary with energy and temperature and differ from the constant value of $\tau_C$ = 10 fs as often assumed in the constant relaxation time approximation (CRTA) \cite{sahni2020reliable,vikram2019accelerated}. The variation of individual scattering processes with energy and temperature are also shown in Fig. S5 and Fig. S6 for p-type and n-type, respectively (Supplementary Information). 

\subsection{Thermoelectric (TE) coefficients}
Next, we present the TE coefficients with respect to the Fermi level's relative position, $E\textsubscript{F}$, which correlates directly with doping density. Here, $E\textsubscript{F}$ is set to 0 eV at the intrinsic Fermi level in the bandgap (typically not the midgap level due to VB/CB asymmetry) while the midgap is at $E\textsubscript{mid}$ = -0.037 eV. The CB minima is placed at $E\textsubscript{C}$ = 0.22 eV whereas the VB maxima at $E\textsubscript{V}$ = -0.29 eV. All the relevant scattering mechanisms such as ADP and ODP (intra and inter-valley), POP and IIS are considered. 
\begin{figure*}[htbp]
	\centering
	
    \includegraphics[scale = 0.7]{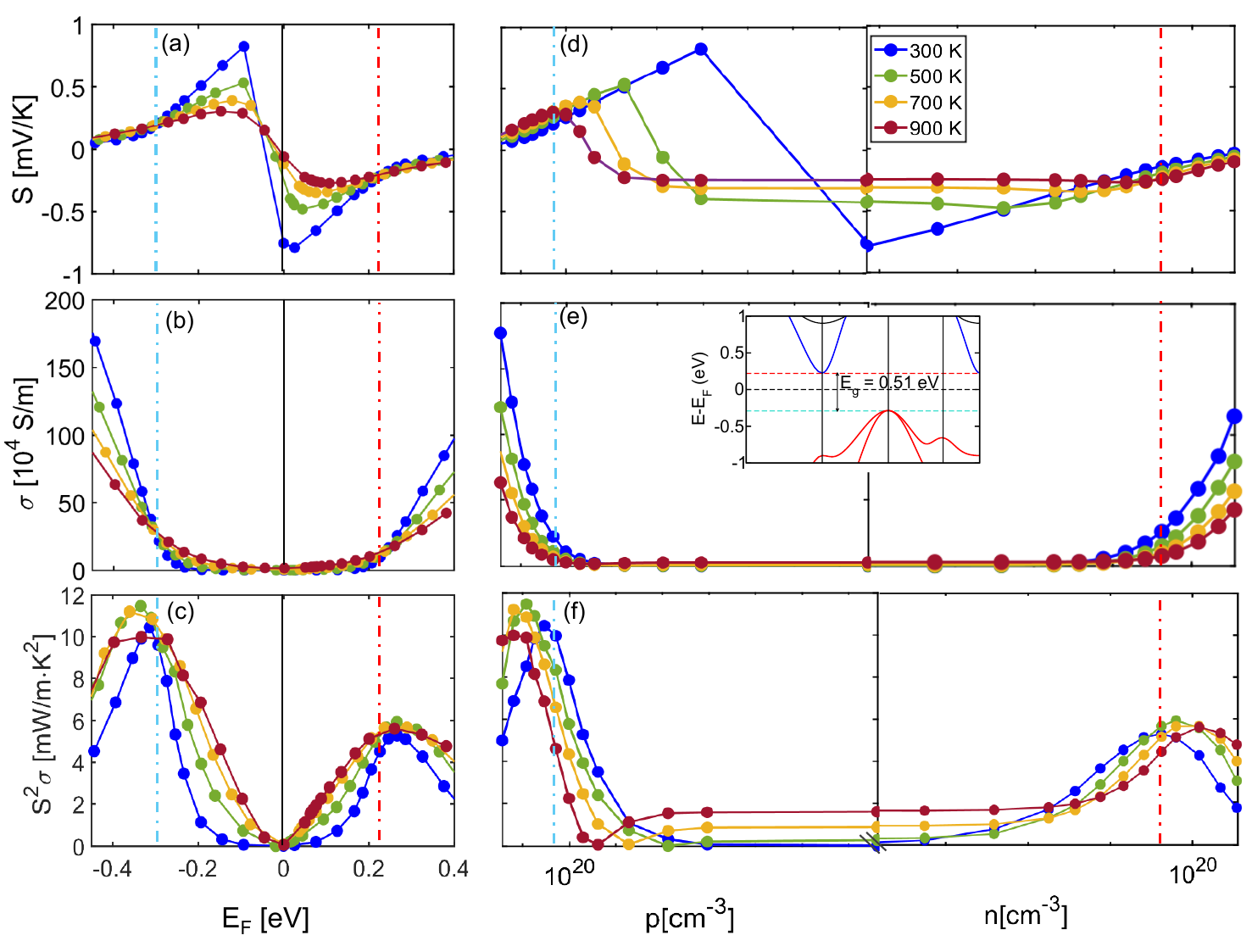}
\caption{Electronic transport properties: (a) Calculated Seebeck coefficient, (b) electrical conductivity and (c) power-factor versus the Fermi level position $E\textsubscript{F}$ (left-panel) and versus the doping density (right-panel). Here we consider all relevant scattering mechanisms (ADP, ODP, POP, IIS). The blue and red dashed lines show the band edges and the black line shows the intrinsic fermi level.}
	\label{TE_plots}
\end{figure*}
\par
 The comprehensive computed electronic transport properties ($S$, $\sigma$ and $\sigma S^2$) versus the relative position of Fermi-level $E\textsubscript{F}$ (left panel), as well as versus the doping density (right-panel) for p-type and n-type NbFeSb for different temperatures, are shown in Fig. \ref{TE_plots}. With the black, blue and red vertical lines, we indicate the position of the intrinsic Fermi level, the VB edge and the CB edge, respectively. Due to the asymmetry between the masses in the VB and CB, the zero crossing of the Seebeck coefficient is shifted from the charge neutrality point towards the VB, as the Seebeck coefficient becomes zero at the point of equal conduction between the VB and the CB (more evident in Fig. \ref{TE_plots}(b)). Note that the effective mass of holes is significantly greater than that of electrons, resulting in lower mobility for holes compared to that for electrons (see Fig. S13 in SI). Thus, the point of equal conductivity between holes and electrons is shifted towards the lower conductivity band (VB) \cite{graziosi2020ultra}. 
 \par
The peak power factor for p-type is obtained to be around \textcolor{black}{10.44 mW/mK$^2$ at 300 K, while it peaks to 11.45 mW/mK$^2$ at 500 K at a carrier concentration of p = 8.4 x 10$^{20}$ cm$^{-3}$. For n-type, the computed peak power-factor is 5.28 mW/mK$^2$ at 300 K, while it peaks to 5.92 mW/mK$^2$ at 900 K at a carrier concentration of n = 6.27 x 10$^{19}$ cm$^{-3}$}.  Thus, the n-type peak can be obtained at a doping density an order of magnitude lower than that for p-type NbFeSb, a consequence of the higher effective mass ($m\textsubscript{DOS}$ of holes as compared to that of electrons, as shown in Fig. \ref{TE_plots}(b).  In general, for n-type the PF peak for the various temperatures is achieved when $E\textsubscript{F}$ is placed somewhat more into the bands compared to p-type (in the case of T = 300 K the n-type PF peak is achieved when $E\textsubscript{F}$ is placed \textcolor{black}{0.04} eV into the CB, whereas for p-type when $E\textsubscript{F}$ is 0.02 eV into the VB).
\par
At high temperatures, the minority carriers can be thermally generated. In the case where the Fermi level is placed in the bandgap, the absence of doping can lead to intrinsic phonon-limited carrier mobility. In addition, the asymmetry between the transport in the VB and CB, which leads to an asymmetric zero Seebeck crossing, can provide finite Seebeck coefficients, which together with phonon–limited conductivity, can lead to high power-factor values around the intrinsic $\eta\textsubscript{F}$ = 0 eV region \cite{graziosi2022bipolar}. As shown in Fig. \ref{TE_plots}(f), the PF remains around \textcolor{black}{1-2 mW/mK$^2$ at 700-900 K} for a large range of carrier concentrations for the case where the Fermi level is in the bandgap (region between the blue dashed lines). Note that much higher values of PFs can be obtained in this low carrier density region under different conditions. As we have shown in Ref. \cite{graziosi2022bipolar}, asymmetric band structures with regards to the CB and VB can allow for very high bipolar PFs compared to their unipolar ones, in materials in which the intrinsic phonon conductivity is high in the bandgap regions where the Seebeck is finite, assisted by the absence of the strong IIS mechanism. 
\begin{figure}[h!]
	\centering
	\includegraphics[scale = 0.7]{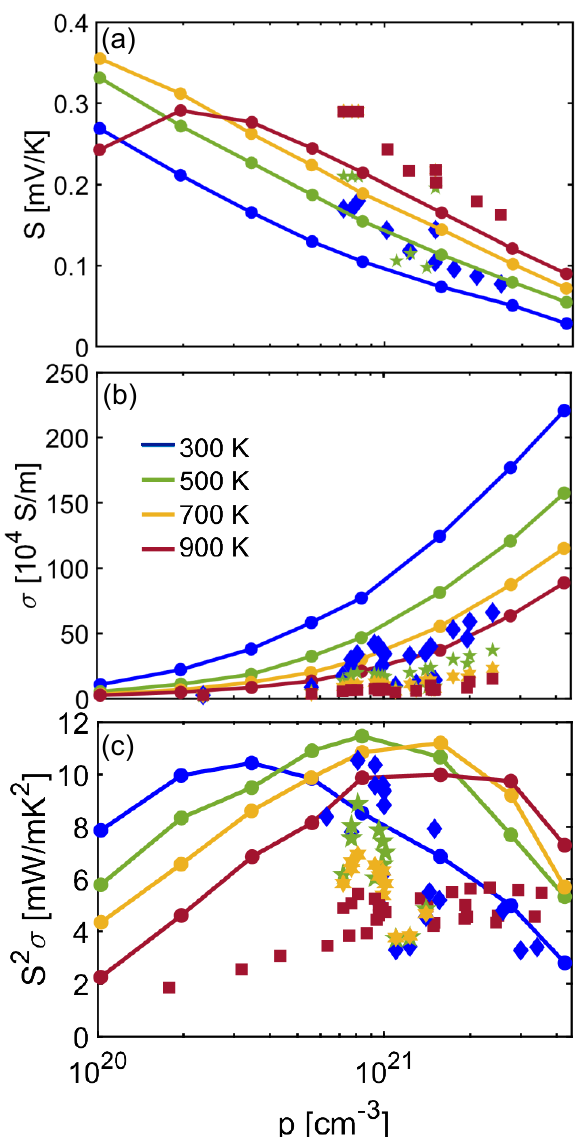}
\caption{Comparison with experimental values: Our calculations for: (a) Seebeck coefficient, (b) electrical conductivity, and (c) power factor for p-type NbFeSb versus doping concentration for various temperatures. The diamond, star, hexagram, and square markers correspond to experimentally reported values from references\cite{fu2014high,yu2018unique,fu2016enhancing,ren2018ultrahigh,joshi2014nbfesb,fu2015band,kahiu2022optimized,fu2015realizing,he2016achieving}}
	\label{comparison}
\end{figure}
\begin{figure*}[ht]
    \centering
    \begin{minipage}{0.7\textwidth}
        \includegraphics[width=\textwidth]{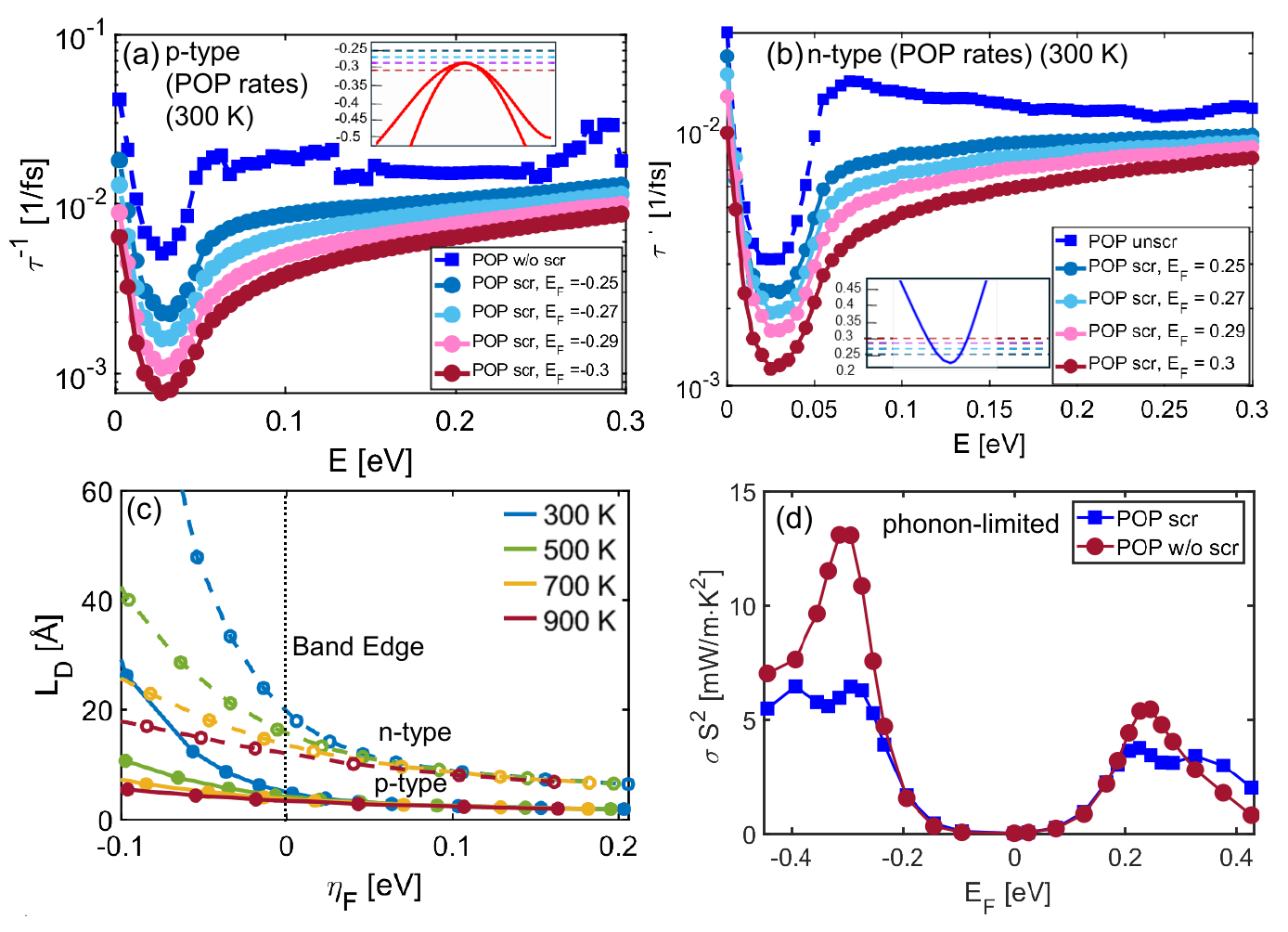}
    \end{minipage}%
    \hfill
    \begin{minipage}{0.25\textwidth}
        \vspace*{\fill} 
\caption{Importance of POP screening: (a) and (b) show the p-type and n-type POP rates at 300 K, which decrease with increasing carrier concentration. (c) The Debye screening length as a function of  $\eta\textsubscript{F}$ at different temperatures for holes (solid lines) and electrons (dashed lines) (the valence and conduction band edges are shifted at the same x-axis point as indicated for comparison. (d) Overall phonon-limited power factor at 300 K for the cases of with and without screening in the treatment of POP.}
	\label{screening}
\vspace*{\fill}
    \end{minipage}
\end{figure*}
\par
With regards to the accuracy of the simulations, in Fig. \ref{comparison}, we compare our results with the previously experimentally reported transport coefficient values for p-type NbFeSb. The power-factor values calculated using our method agree well with previous experimental reports. Most of the reported data points are in the region where the optimal carrier concentration is obtained. In most experimental reports, the peak PF value is around PF \textcolor{black}{= 2 - 10.5 mW/mK$^2$ in the temperature range from 400-900 K and the carrier concentration is around p = 6.3$\times$ 10$^{20}$ - 1.5 $\times$ 10$^{21}$ cm$^{-3}$. Our peak PF of around 11.5 mW/mK$^2$ at 500 K at carrier concentration of 8.4 $\times$ 10$^{20}$ cm$^{-3}$ is somewhat higher. However, it is expected that the theoretical values are generally higher than the experimental values, because of various factors such as the presence of defects, dislocations, alloying etc. in the measured structures, which are not accounted for in simulations. These factors almost certainly reduce the electrical conductivity and mobility significantly. Overall, the electrical conductivity of the experimental data is around a factor of three lower compared to our computed results, as shown in Fig.\ref{comparison}(b). On the other hand, the measured Seebeck coefficient is larger compared to the computed Seebeck coefficient, following the well-expected adverse trend compared to the electrical conductivity, as shown in Fig. \ref{comparison}(a). The electrical conductivity suffers much more in experiment compared to the improvements in the Seebeck coefficient (compared to the simulated data), such that the PF is overall lower the experiment (especially for higher temperatures)}.

\subsection{The effect of screening}
An important consideration for POP and IIS is the effect of screening. In the case of IIS, the free charge carriers can shield the dopant ions and mitigate the scattering rates. Similarly, screening also weakens the POP scattering mechanism and needs to be accounted for. In this case the scattering rates are reduced due to screening of the POP dipole electric field by free carriers. Considering screening is computationally much more expensive, because the scattering rates are not only energy dependent alone, but are also Fermi level dependent. Thus, they need to be computed separately for every $E\textsubscript{F}$ position. Screening considerations, however, are quite important to accurately compute the electronic transport properties.
\par
In Fig. \ref{screening}(a) and \ref{screening}(b) we show the \textcolor{black}{POP rates up to $\eta\textsubscript{F}$ values of -0.3 and 0.3 for p-type and n-type respectively, covering the region where the power-factors peak in each case (see insets for the actual positions of the Fermi level)}. The POP scattering rates for both p-type and n-type decrease with the increase in carrier concentration, a consequence of increased screening. The POP rates for the unscreened case (blue lines) are larger for p-type as compared to n-type. This is due to the presence of two degenerate valence bands, which introduces inter-band scattering, in addition to the intra-band scattering between the heavy mass p-type bands (see Eq. 9 in methods). In fact, a single heavier band mass (as in the VB) will result in larger exchange vectors and reduce the POP and IIS rates. However, having a second heavier band overlapping at the same k-space region, enables small exchange vector transitions between them (from geometrical considerations), which increases the scattering rates (and makes the overall exchange vector smaller - note that inter-valley transitions involve large exchange vectors, but these scattering rates are small and do not affect the overall scattering significantly). When screening is considered, the p-type POP rates decrease more with increasing carrier concentration as compared to the n-type rates. The reason behind this behavior can be explained by examining the generalized screening term for POP rates, defined as  $\left[\frac{\textbf{q}^2}{(\textbf{q}^2+\frac{1}{L_D^2})}\right]^2$. Here \textbf{q} is the momentum exchange vector, $L\textsubscript{D} = \sqrt{\frac{k_\infty \epsilon_0}{e} \left( \frac{\partial n}{\partial E_F} \right)^{-1}}$
 is the screening length, $E\textsubscript{F}$ is the Fermi level and \textit{n} is the carrier density. The screening length $L\textsubscript{D}$ for holes and electrons is shown by the solid and dashed lines, respectively, for different temperatures in Fig. \ref{screening}(c) at the same reduced Fermi levels (the band edge for both cases is noted and placed at $\eta\textsubscript{F}$ = 0 eV). In general, it decreases with $E\textsubscript{F}$ (and density) indicating reduction in the scattering rates, and it is lower for holes compared to electrons (solid lines are lower compared to the dashed lines), indicating lower scattering rates for holes (note that low $L\textsubscript{D}$ means that the scattering interaction is screened at very small length and the scattering rate is thus low). The reason for this behavior is the larger density of states (DOS) for holes, which reduces $L\textsubscript{D}$ since it involves the $(\frac{\partial n}{\partial E_F})^{-1}$ term, which decreases as the DOS increases. Thus, around the energy regions at which the PF peaks, the holes scattering rates are becoming weaker, approaching similar values as those for electrons. \textcolor{black}{Note that the screening length under simplifications is proportional to $\sqrt{T/n}$. Here, since we compare at the same $\eta_F$, the density increases with temperature as well (more than linearly), which makes the $L_D$ seem to reduce with T, rather than increase.}
\par
The phonon-limited PF values with and without considering POP screening show large differences especially for p-type transport, as shown in Fig. \ref{screening}(d). This signifies that screening needs to be considered, despite the fact that the entire computation is now in addition Fermi level dependent. This is more important with materials with large density of states and elongated, dispersive bands (p-type here), rather than materials with smaller effective masses and narrow bands (n-type here). We stress, however, this is specifically the case for NbFeSb. In our calculations for different half-Heusler materials, these conclusions could vary. 

\subsection{Intra- versus inter-valley transition strength}
It is also important to highlight the significance of intra- versus inter-valley transitions, since this is crucial information for studying band alignment in TE materials \cite{kumarasinghe2019band,akhtar2025conditions}. For this, the scattering rates from Fig. \ref{intra_vs_inter} were averaged in the energy range of $E$ = 0 - 0.1 eV, to provide an estimate of the relative strength of these processes. The deformation potential interaction can be both intra-valley and inter-valley, whereas the polar optical scattering is mostly intra-valley because of its long-range nature (|\textbf{q}|$\textsuperscript{-1}$). The inter-valley scattering is essentially the IVS non-polar optical processes, whose strength is very small (shown in red in Fig. \ref{intra_vs_inter} (a, b)) for both the p-type and n-type NbFeSb cases. Thus, the intra-valley processes dominate in this material. We expect this to be the case broadly in the family of polar half-Heusler materials, but also in other polar TE materials as is in the case of Mg$_3$Sb$_2$ as well\cite{li2024efficient}. Note that it is not just inter-valley scattering that is detrimental for band alignment strategies, but also intra-band scattering, which can be strong in the case where the valleys of the aligned and base bands are at the same position in the Brillouin zone (as in the VB of NbFeSb here, although these two bands are actually aligned) \cite{park2021band}.  

\subsection{\textcolor{black}{Influence of scattering mechanisms in transport}}
\begin{figure*}[ht]
    \centering
    \begin{minipage}{0.7\textwidth}
        \includegraphics[width=\textwidth]{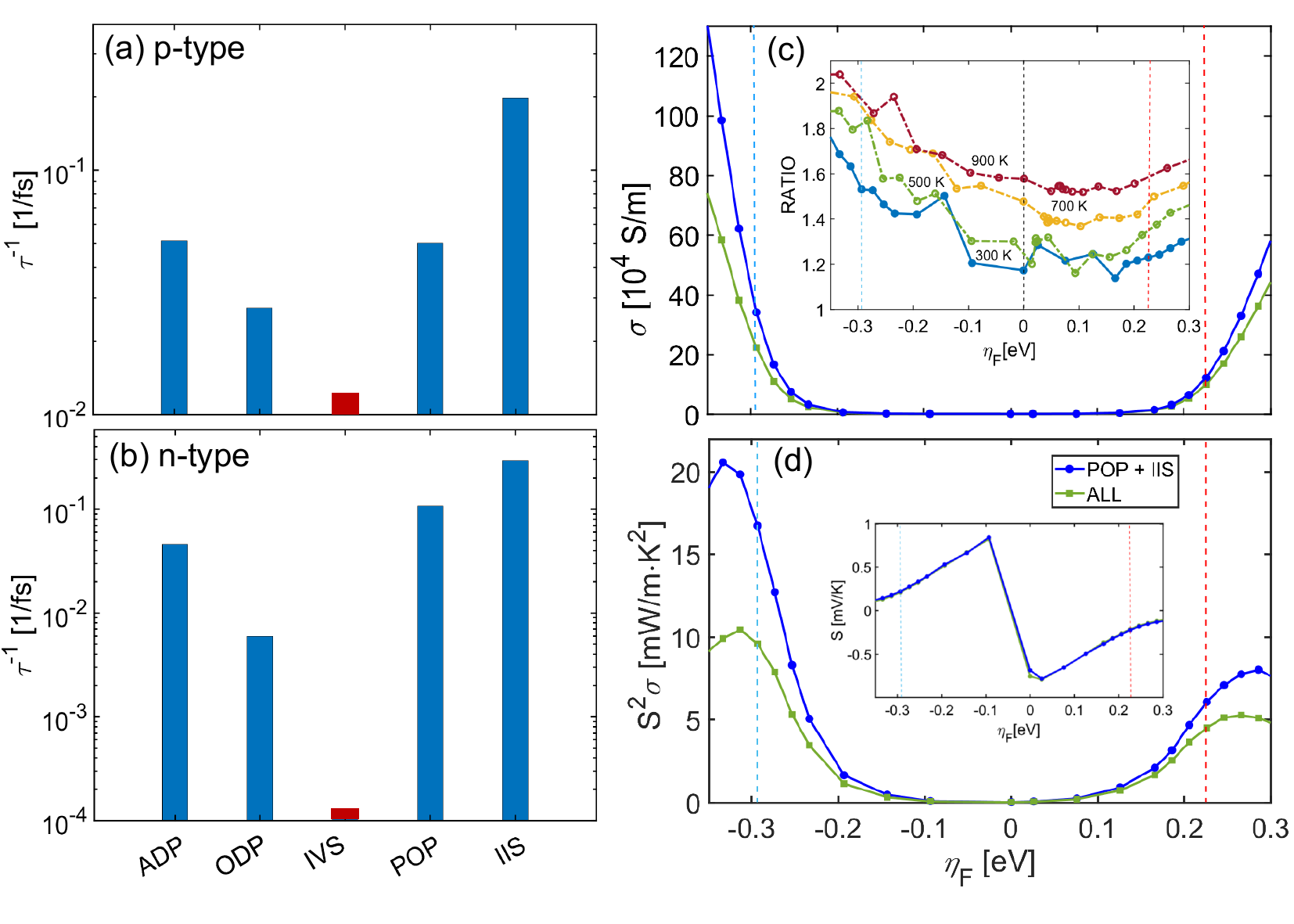}
    \end{minipage}%
    \hfill
    \begin{minipage}{0.25\textwidth}
        \vspace*{\fill} 
\caption{Scattering strength comparison at 300 K: (a-b) Comparison between the scattering strengths of intra and intervalley scattering mechanisms for p-type and n-type. \textcolor{black}{(c) Comparison between the conductivity and (d) power-factor obtained by including all scattering mechanisms versus that obtained by including only POP and IIS, which are the two dominant scattering mechanisms. The inset in (c) shows the ratio of $\sigma$ obtained by including only POP + IIS vs ALL scattering mechanisms. The inset in (d) shows the comparison between the Seebeck obtained by including all scattering mechanisms versus that obtained by including only POP and IIS.}}
	\label{intra_vs_inter}
\vspace*{\fill}
    \end{minipage}
\end{figure*}

\textcolor{black}{Since IIS in p-type and POP and IIS in n-type are the strongest scattering mechanisms, we now compute how much of the PF these mechanisms determine. For this, we compute the PF considering only the POP and IIS scattering mechanisms and compare this with the overall PF values (considering all scattering mechanisms). The two calculations are shown in Fig. \ref{intra_vs_inter}(c) and (d) for $\sigma$ and $S^2\sigma$, respectively. The ratio of the $\sigma$ values taking into account POP+IIS to those calculated by taking into account all scattering mechanisms is shown in the inset of Fig. \ref{intra_vs_inter}(c). Using resistance combination consideration, a ratio of 2 signifies equal strength between IIS + POP and the non-polar ADP + ODP, while a ratio of 1 shows that IIS + POP has full dominance over ADP + ODP. At 300 K (blue line), this ratio is around 1.5 and 1.2 at the valence and conduction band edges, respectively. This indicates that the POP + IIS scattering mechanisms are around twice as strong in determining $\sigma$ compared to ADP + ODP for p-type, and five times as strong for n-type (again using resistance combination considerations). For p-type the discrepancy is somewhat larger due to the fact that the non-polar scattering rates are higher compared to the rates in the n-type case (see Fig. \ref{rates_tau}(a), and also the smaller screening lengths in the heavier VB, which weaken the POP and IIS rates). Figure \ref{intra_vs_inter}(d) shows that the conductivity differences are almost entirely transferred to the PF, as the Seebeck coefficient (inset of Fig. \ref{intra_vs_inter}(d)) is almost identical between the two simulation cases (at 300 K).} 
\par
\textcolor{black}{Note, however, that this comparison is shown for 300 K. At higher temperatures, the comparative strengths would change in favor of the non-polar ADP + ODP mechanisms - our calculations show that the influence of POP + IIS scattering compared to ADP + ODP in determining $\sigma$ is approximately halved (compared to their influence at 300 K) with doubling the temperature. In general, the ratio in the inset of Fig. \ref{intra_vs_inter}(c) increases with T. In the case of p-type, at 900 K, the strength of the Coulombic and non-polar mechanisms becomes equal. For n-type, even at high T, the Coulombic IIS + POP are still much stronger. Thus, even in polar materials such as this one, accurate computation of the non-polar scattering component could be important for the PF, especially for high T. An inaccurate computation of the acoustic and optical deformation potential scattering can lead to significant quantitative error in the transport properties. }

\subsection{Computational Performance}
In this subsection, we highlight the major computational efficiencies provided by our approach, specifically compared to the full DFPT + Wannier methods, as for example in EPW \cite{lee2023electron}. The overall computation can be split into two steps. The first involves calculating electron-phonon matrix elements via the DFT and DFPT methods. The second pertains to the transport computation. The first step, which identifies the transitions and extracts the selected matrix elements for those transitions is the same in our case as in the DFPT + Wannier methods. For the first step, we used a k-mesh of 12 $\times$ 12 $\times$ 12 for DFT and a q-mesh of 6 $\times$ 6 $\times$ 6 for DFPT. This step is the most time-consuming part of the calculation, requiring around 12,000 CPU hours.
\par
The second step, involving the transport calculation, is computationally significantly less expensive in comparison to DFPT + Wannier based approaches. Full DFPT + Wannier approaches require a fairly dense post-interpolation mesh and the consideration of millions of matrix elements contributing to transport, in order to obtain convergent results \cite{ponce2018towards}. This could lead to computational challenges, such as the requirement of higher memory nodes as well as more core hours, especially in case of complex systems with large unit cells and low symmetry \cite{li2024efficient}. In our method, the transport simulation step requires only 6,300 CPU hours for both electrons and holes which is significantly lower compared to what DFPT + Wannier methods demand. The transport computation from DFPT + Wannier method is expected to require around 70,000 CPU hours or even more, depending on convergence, for only one type of carrier, as elaborated on in our previous works\cite{li2024efficient}. Thus, effectively, we can reduce the overall computational cost by around 20 times for full bipolar transport using our approach compared to the full DFPT + Wannier method. \cite{li2021deformation,li2024efficient} \textcolor{black}{(We would like to note that we tried substantially to simulate this material (NbFeSb) using EPW in order to compare with our results. However, achieving quantitative convergence with EPW proved very difficult due to the extremely fine Brillouin-zone sampling needed).} In ElecTra, although we extract just a limited number of matrix elements and converting them into deformation potentials, we intentionally concentrate on the critical transport areas, leading to a locally concentrated grid of matrix elements. This approach might be more beneficial than choosing matrix elements on a sparse grid throughout the entire Brillouin zone \cite{lee2023electron}. More developments in this area could lead to considerable reduction in computational costs making it feasible to compute accurate deformation potentials for large number of systems and even bigger with respect to unit cell size and complexity. \textcolor{black}{We also performed AMSET\cite{ganose2021efficient} calculations for NbFeSb and compared the result to our ElecTra computations. We found that for this material, the two results differ in some cases, with AMSET predicting higher power factor compared to the ElecTra results. AMSET derives a global deformation potential value for the non-polar electron-phonon scattering contributions based on the band shifts when straining the material, and thus does not make distinction between ADP and non-polar ODP scattering. However, in certain cases, for example if we exclude the non-polar optical deformation potential scattering and screening due to POP, the values obtained from AMSET and ElecTra are in very good agreement.} 

\section{Conclusions}
In summary, we have studied with \textit{ab initio} simulations the thermoelectric transport properties and the role of the different scattering processes in the high-performance thermoelectric (TE) material NbFeSb. \textcolor{black}{Our computation shows that at 300 K, the POP + IIS scattering mechanisms are around twice as strong in determining $\sigma$ and the PF compared to ADP + ODP for p-type carriers, while for n-type carrier POP + IIS is five times stronger compared to scattering from the non-polar ADP + ODP mechanisms, although these numbers reduce as the temperature increases.} Our calculations also suggest that the intra-valley processes dominate in this material, owing to the strength of IIS, POP, ADP, ODP over IVS (see Fig. \ref{intra_vs_inter}). Screening weakens POP and IIS for the larger DOS valence band more compared to the lower DOS conduction band, but they still remain strong to determine the PF (especially for n-type). \textcolor{black}{We show that the peak PF for p-type is 11.45 mW/mK$^2$ with very good agreement with multiple experiments. We also show that the peak PF for the very less explored experimentally n-type case, is lower at 5.9 mW/mK$^2$. For n-type the PF peak is reached at densities around n = 6.27 $\times$ 10$^{19}$ cm$^{-3}$, which are almost an order of magnitude lower compared to where the peak of the PF in the p-type case is reached. Still, however, although the calculated PF values are somewhat close with the experimental values, we show that the electrical conductivity in experiment is three times lower compared to our calculations, while the experimental Seeebeck coefficients are somewhat higher. This suggests that better crystallinity and reduced defects could allow even up to doubling the PFs in certain cases.} This information offers guidance of performance optimization through selection of appropriate doping levels. 
\par
Regarding our method, we have used an approach which involves the extraction of deformation potentials $\textit{ab initio}$, to calculate scattering rates to be used in a BTE simulator. Our method uses a limited number of matrix elements near specific \textbf{q}-vectors related to intra/inter-valley scattering, which can be chosen on a fine mesh. Crucially, by considering each \textbf{q}-vector and phonon branch individually, we gain insight into the underlying fundamental physical processes, including intra- versus inter-valley transitions and the strengths of each mechanism separately. Our approach achieves ab initio accuracy, at significant computational cost reductions compared to fully ab initio DFPT + Wannier techniques by at least 20 times due to the significantly fewer matrix elements needed. Thus, we believe that our approach can be extremely useful for understanding transport properties, but can also be scalable to provide high quality training data for further machine learning related research. 

\section{Methods}
The electronic band structure, phonon dispersion, and electron-phonon coupling matrix elements are determined using DFT and DFPT through the Quantum ESPRESSO package \cite{giannozzi2009quantum}. The calculations employ optimized norm-conserving Vanderbilt (ONCV) pseudopotentials within the framework of the generalized gradient approximation (GGA) utilizing the Perdew-Burke-Ernzerhof (PBE) functional. The EPW package \cite{lee2023electron} is utilized to carry out Wannier function interpolation for the electron-phonon coupling matrix elements.
\par
For transport calculations for NbFeSb (having 3 atoms/unit cell), a k-mesh of 81 $\times$ 81 $\times$ 81 was used. We use our BTE simulator ElecTra for calculating the transport properties \cite{graziosi2023electra}. 
The scattering rate expressions for different scattering processes used are as follows.
          For ADP: 
	\begin{equation}
|S_{\textbf{k},\textbf{k'}}^{\text{ADP}}| = \frac{\pi}{\hbar} D_{\text{ADP}}^2 \frac{k_B T}{\rho v_s^2} g(E)
	\end{equation}                  
where $\rho$ is the mass density and $g$(E) is the density of states of the final scattering state. 
           For ODP: 
	\begin{equation}
|S_{\textbf{k},\textbf{k'}}^{\text{ODP}}| = \frac{\pi D_{\text{ODP}}^2}{2 \rho \omega} \left(N_\omega + \frac{1}{2} \mp \frac{1}{2} \right) g(E \pm \hbar \omega)
    \end{equation}
where $\omega$ is the dominant frequency of the optical phonons, considered to be constant over the entire reciprocal unit cell. $N_\omega$ is the phonon Bose-Einstein statistical distribution and the + and - signs indicate the emission and absorption processes, respectively.
For IVS, a similar expression as ODP is used:
	\begin{equation}
|S_{\textbf{k},\textbf{k'}}^{\text{IVS}}| = \frac{\pi D_{\text{IVS}}^2}{2 \rho \omega} \left(N_\omega + \frac{1}{2} \mp \frac{1}{2} \right) g(E \pm \hbar \omega)
   \end{equation}                     
where $\omega$ is the phonon frequency associated with the corresponding inter-valley scattering processes.
\textcolor{black}{For POP, we use the Fr\"ohlich formalism as: 
	\begin{equation}
    \begin{split}
|S_{\textbf{k},\textbf{k'}}^{\text{POP}}| = \frac{\pi e^2 \omega}{|k-k'|^2 \epsilon_0} \left( \frac{1}{k_\infty} - \frac{1}{k_0} \right) \left( N_\omega + \frac{1}{2} \mp \frac{1}{2} \right) \times \\ 
 g(E \pm \hbar \omega) <I_{\mathbf{k,k'}}^2>
 \end{split}
	\end{equation}
where $e$ is the electronic charge, and $\omega$ is the dominant frequency of polar optical phonons over the whole Brillouin zone, which has been validated to be a satisfactory approximation\cite{li2024efficient}. $k_0$ is the static dielectric constant and $k_\infty$ is the high-frequency dielectric constant.} \textcolor{black}{Here, $I_{\textbf{k,k'}}$ is the overlap integral which is given by: 
\begin{equation}
I_{mn}(\mathbf{k},\mathbf{k}+\mathbf{q})
\;=\;
\bigl\langle u_{m,\mathbf{k}+\mathbf{q}} \mid u_{n,\mathbf{k}}\bigr\rangle,
\end{equation}
and is computed directly from DFT wave functions stored in the Quantum ESPRESSO output. To account for overlaps around the band extrema efficiently, 100 randomized k-points are placed within a 10\% reciprocal lattice vector radius. We then compute $\sim$\textit{N}$^2$ intra-band and inter-band overlap pairs by considering all possible initial and final k-state transitions among these points. The values for both intra-and inter-band transitions in the VB vary significantly in the range from 0 to 1 with an average of 0.64, and an average of the overlap squared values, that enter the scattering rates, around 0.5. Detailed explanation of overlap integrals is provided in the SI file.}
For IIS, we use the Brooks-Herring model as: 
	\begin{equation}
|S_{\textbf{k},\textbf{k'}}^{\text{IIS}}| = \frac{2\pi}{\hbar} \frac{Z^2 e^4}{k_0^2 \epsilon_0^2} \frac{N_\text{imp}}{|\textbf{k}-\textbf{k'}|^2 + \frac{1}{L_D^2}} g(E)<I_{\mathbf{k,k'}}^2>
    \end{equation}                                                  
where $Z$ is the electric charge of the ionized impurity (we assume $Z$ = 1 here), $N_{imp}$ is the density of the ionized impurities, and $L\textsubscript{D}$ is the generalized screening length given by $L\textsubscript{D} = \sqrt{\frac{k_s \epsilon_0}{e} \left( \frac{\partial n}{\partial E_F} \right)^{-1}}$, where $E\textsubscript{F}$ is the Fermi-level and \textit{n} is the carrier density. 
The overall transition rate,$|S_{\textbf{k},\textbf{k'}}|$, is calculated by combining the strength of all scattering mechanisms using Matthiessen’s rule as:
\begin{equation}
|S_{\textbf{k},\textbf{k'}}| = |S_{\textbf{k},\textbf{k'}}^{\text{ADP}}| + |S_{\textbf{k},\textbf{k'}}^{\text{ODP}}| + |S_{\textbf{k},\textbf{k'}}^{\text{IVS}}| + |S_{\textbf{k},\textbf{k'}}^{\text{POP}}| + |S_{\textbf{k},\textbf{k'}}^{\text{IIS}}|
\end{equation}

\textcolor{black}{\indent After calculating the scattering rates, the thermoelectric coefficients, namely  the electrical conductivity, $\sigma$, and Seebeck coefficient, $S$ are calculated as:
 \begin{equation}
     \sigma_{ij} = e^2 \int_E \Xi_{ij} (E) \Big(\frac{\partial f_0}{\partial E} \Big) dE 
 \end{equation}
  \begin{equation}
     S_{ij} = \frac{e k_{\text{B}}}{\sigma_{ij}} \int_E \Xi_{ij} (E) \Big(\frac{\partial f_0}{\partial E} \Big) \frac{E - E_\text{F}}{k_\text{B} T} dE 
 \end{equation}}

\textcolor{black}{Here, $f_0$ is the equilibrium Fermi-Dirac distribution function, $\Xi_{ij} (E)$ is transport distribution function (TDF) defined as:
\begin{equation}
    \Xi_{ij} (E) = \int_E \tau_{{\textbf{k},\textbf{k}'}}(E) v_{ij}^2(E) g(E)
\end{equation}
where, $\tau_{\textbf{k},\textbf{k}'}$ is the relaxation time (inversely proportional to the transition rates), $v(E)$ is the bandstructure velocity, $g(E)$ stands for the density of states at energy $E$, and $i,j$ are the Cartesian coordinate indexes, for which we set $i=j=x$).}

\section*{Author contributions}
BS, ZL and NN conceived the project. BS performed the calculations, analyzed the results and drafted the manuscript. BS and YZ computed interband and intraband overlap integrals. NN supervised the work, analysed the results and helped in drafting the manuscript. PG helped in TE calculations using the Electra code. YZ, ZL, PG and RD helped in editing and finalizing the draft.

\section*{Conflicts of interest}
There are no conflicts to declare.

\section*{Data availability}
The ElecTra code (for calculating thermoelectric properties) can be found at: https://github.com/PatrizioGraziosi/ELECTRA ElecTra (v1.4) was used for calculations. The EMAF code (for computing effective masses) can be found at: https://github.com/PatrizioGraziosi/EMAF-code. Quantum espresso (v6.8) was used for DFT, DFPT and matrix elements calculations. The input files and codes for calculating deformation potentials can be found at https://github.com/Computational-Nanotechnology-Lab/Deformation-potential-extraction. Part of the data supporting this article has been included in the SI. Larger data files can be provided upon request. Supplementary information: Atom-projected electron and phonon density of states, intravalley electron–phonon matrix elements for electrons and holes in different crystallographic directions, overlap integrals calculation formalism BTE input parameters, variation of electron–phonon scattering rates with temperature and mobility of electrons and holes at different temperatures. See DOI: https://doi.org/10.1039/d5mh00228a
                                      
\section*{Acknowledgements}

This work has received funding from the UK Research and Innovation fund (project reference EP/X02346X/1).

\newpage
\section{Supplementary Information}
\subsection{Electronic band structure and density of states }
Here we show the electronic band structure and atom-projected density of states of NbFeSb.  The conduction bands show major contribution from Nb d-orbitals only, while the valence bands show major contribution from Fe d-orbitals, followed by Nb d-orbitals and Sb p-orbitals. There is a hybridization between Nb d-orbitals and Sb p-orbitals in the energy range of 0 to -1 eV. The valence band maximum is located at the L high symmetry point and is doubly degenerate, while the conduction band minimum is at the X high symmetry point. 

 \begin{figure}[htbp!]
	\centering
	\includegraphics[width = \columnwidth]{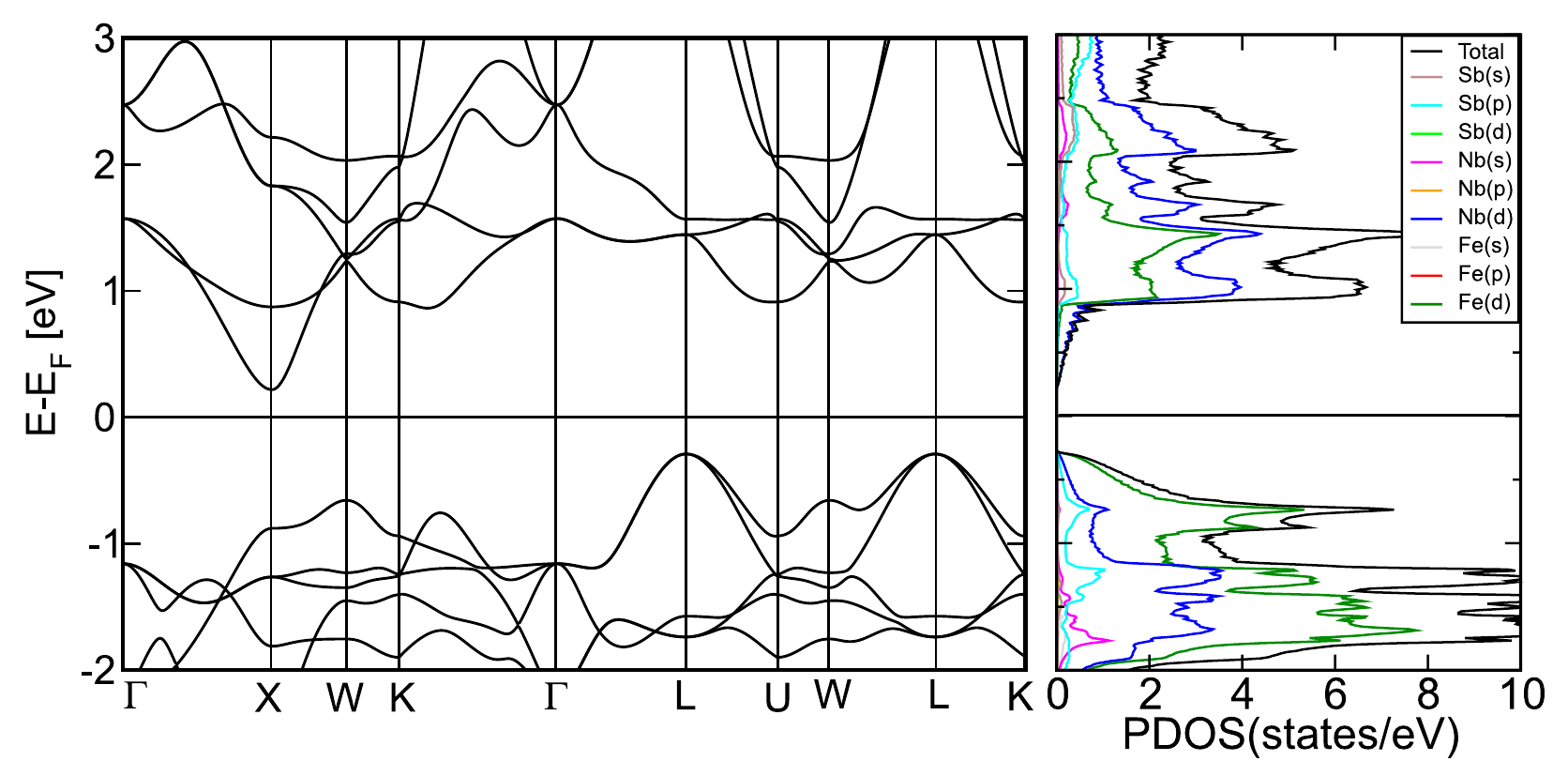}
\caption{Band structure and the atom-projected density of states for NbFeSb}
	\label{S1}
\end{figure}

\subsection{Phonon spectrum and density of states}
\begin{figure}[htbp!]
	\centering
	\includegraphics[width = \columnwidth]{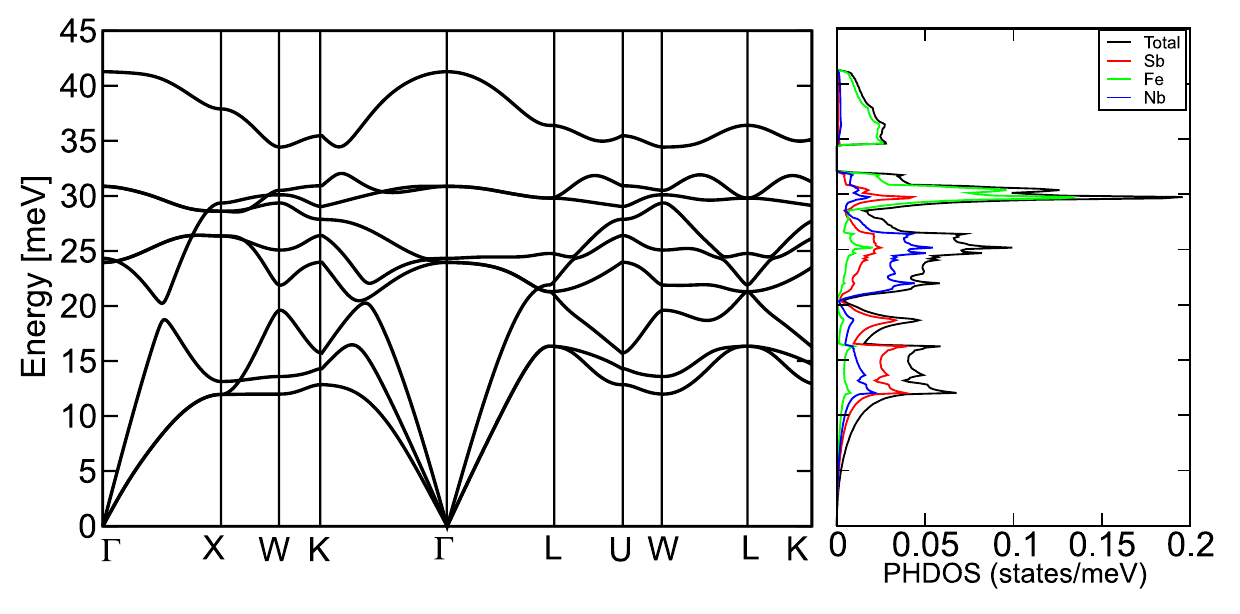}
\caption{Phonon spectrum and the atom-projected phonon density of states for NbFeSb}
	\label{S2}
\end{figure}

 The atom-projected phonon density of states is shown in Fig. \ref{S2} which shows the major contribution of Sb atoms in low frequency regions, followed by contributions from Nb and Fe atoms. The high frequency (longitudinal optical phonon) modes show dominant contribution from Fe atoms.

\subsection{Deformation potentials}
\begin{figure}[h!]
  \centering
  \includegraphics[width = 0.9\columnwidth]{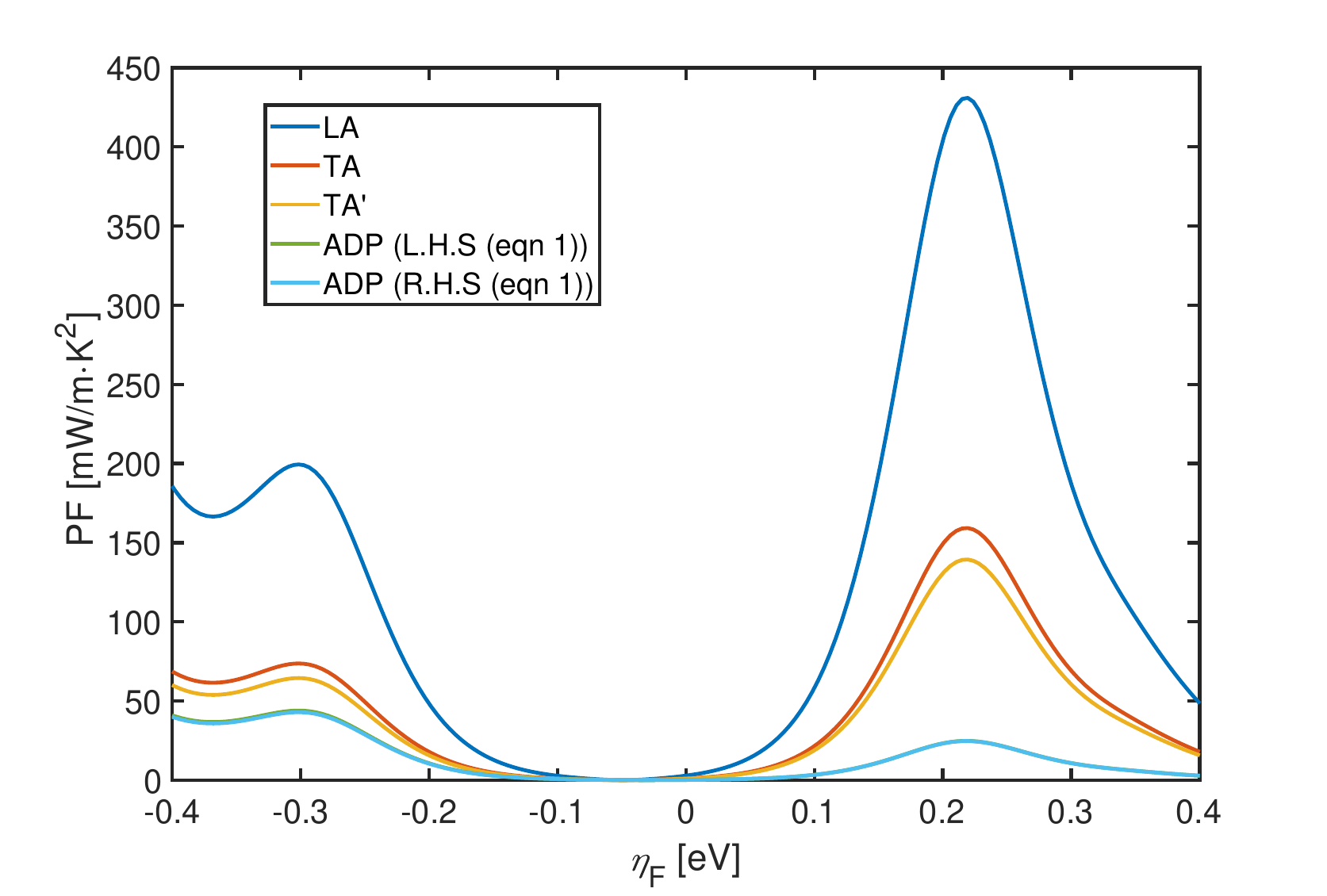} 
  \caption{The PF for the conduction (electrons) and valence (holes) bands, with i) transport limited by LA scattering alone, ii) the two transverse modes alone (TA and TA'), iii) the combined contribution from all three modes (when computed separately) and iv) the combined contribution of the three modes when averaged (as in the paper). In both the electron and hole cases, the latter two results are identical.}
  \label{def}
\end{figure}

\textcolor{black}{While our code can take separate inputs from all phonon branches (acoustic and optical), we find it much easier to reasonably average the values and use a single deformation potential value. We have checked and the results for the power factor are almost identical in the treatment of the average deformation potential, or the separate deformation potentials for each branch. In fact, the average treatment we employ, at least for acoustic phonons, is almost equivalent to treating the different branches separately, i.e.: 
\[
\frac{D_{\mathrm{avg}}^2}{v_{s}^2}
\;=\;
\frac{D_{L}^2}{v_{L}^2}
\;+\;
\frac{D_{T1}^2}{v_{T1}^2}
\;+\;
\frac{D_{T2}^2}{v_{T2}^2}\,.
\]
Each of the terms in the right-hand side of the equation above would enter the scattering rates in an additive way as indicated. Taking the average of the squares in this way is similar to taking branches separately (the sound velocity is not averaged with the squares of the individual velocities, but the difference is accounted for in the average deformation potential extracted).  The figure \ref{def} shows the PF for the full energy scale that spans the conduction (electrons) and valence (holes) bands, as determined by considerations of LA scattering alone, the two transverse modes alone (TA and TA'), the combined contribution from all three modes (when computed separately) and the combined contribution of the three modes when averaged (as in the paper). In both the electron and hole cases, the latter two results are indeed identical.}   

\begin{table}[h]
\centering
\begin{tabular}{|c|c|c|c|c|}
\hline
    \multicolumn{5}{|c|}{Holes}   \\
    \hline
Process &D's  &$\Gamma$-L & $\Gamma$-X &  $\Gamma$-K  \\
\hline
$B1\rightarrow B1$	&ADP (eV)	&2.24	&2.89	&3.84	 \\
	&ODP (eV$\AA^{-1}$)	&2.34	&1.88	&3.26	 \\
\hline
$B1 \rightarrow B2$	&ADP (eV)	&2.53	&1.44	&1.69	 \\
	&ODP (eV$\AA^{-1}$)	&2.82	&2.38	&2.96	 \\
\hline
$B2\rightarrow B1$	&ADP (eV)	&2.24	&2.89	&3.84	 \\
	&ODP (eV$\AA^{-1}$)	&2.34	&1.88	&3.26	 \\
\hline
$B2\rightarrow B2$	&ADP (eV)	&2.53	&1.44	&1.69	 \\
	&ODP (eV$\AA^{-1}$)	&2.82	&2.38	&2.96	 \\
\hline
\end{tabular}
\caption{The intravalley deformation potentials for all four scattering processes (between the two valence bands, B1 and B2) along different crystallographic high-symmetry directions.}
\label{D_process}
\end{table}

\begin{figure}
    \centering
    \includegraphics[width=0.75\linewidth]{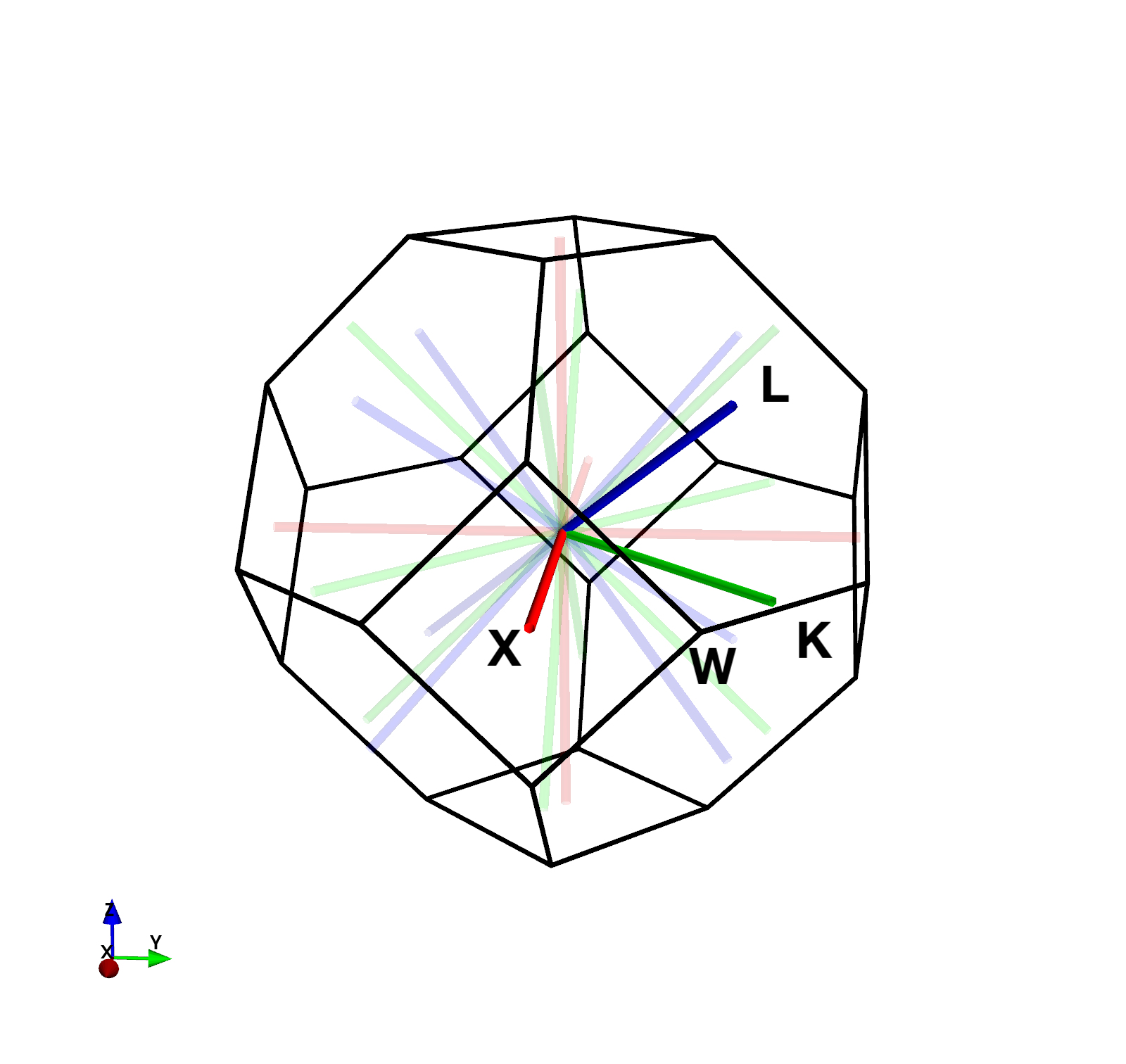}
    \caption{Illustration of high symmetry crystallographic directions, <100> (X), <110> (K), and <111> (L), in the Brillouin zone of a face-centred cubic cell. Symmetrically equivalent directions are indicated by the same colours.}
    \label{fig:bz_dp}
\end{figure}

Directions that are used in the sampling of the deformation potentials are indicated in the Figure \ref{fig:bz_dp}.
Since NbFeSb is a cubic crystal, we can limit the sampling of deformation potentials in high crystallographic directions <100>, <110>, and <111>, (directions with Miller indices with 1 and 0 only) which are equivalent to path from $\Gamma$ to $X$, $K$, and $L$ respectively.
Note, that directions <210> ($W$) is not included in the sampling, as it is not one of those high symmetric crystallographic directions.

\begin{figure*}[htbp!]
	\centering
	\includegraphics[width=1.0\textwidth]{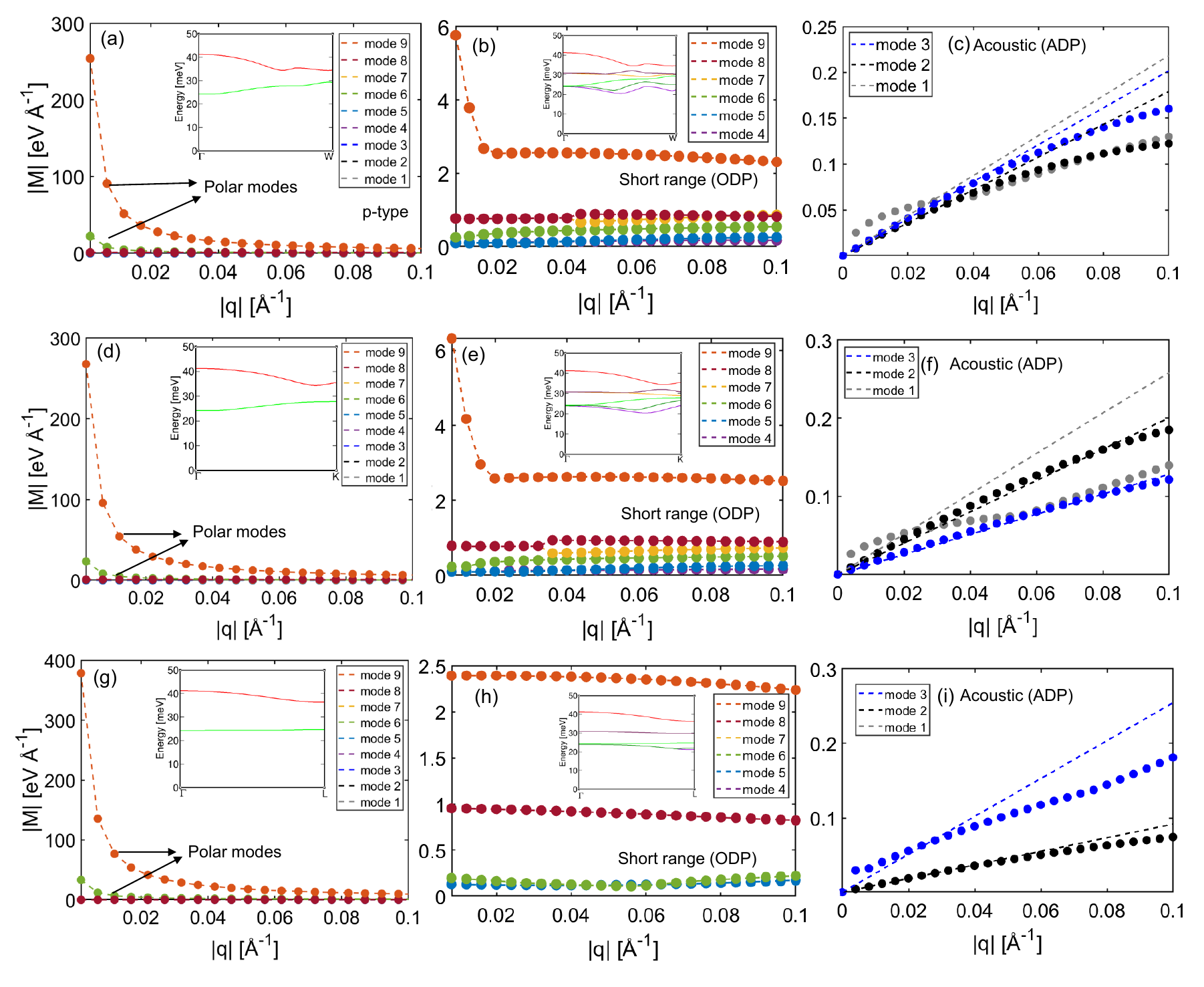}
\caption{Intravalley electron-phonon coupling matrix elements for holes: The matrix elements in various crystallographic directions for B1 $\rightarrow$ B1 along: (a-c)  $\Gamma$- W; (d-f) $\Gamma$- K; and (g-i)  $\Gamma$- L  crystallographic directions. Left column (a, d, g): For all phonons (two modes show polar behaviour) along the $\Gamma$-X line. Middle column (b, e, h): Short range part of matrix elements for optical modes. Right column (c, f, i): acoustic phonon matrix elements for scattering within the valence band maxima (VBM to VBM). The insets in (a, b, d, e, g, h) show the frequencies corresponding to the polar optical and non-polar optical modes respectively.}
	\label{S3}
\end{figure*}

\begin{figure*}[htbp!]
	\centering
	\includegraphics[scale=0.5]{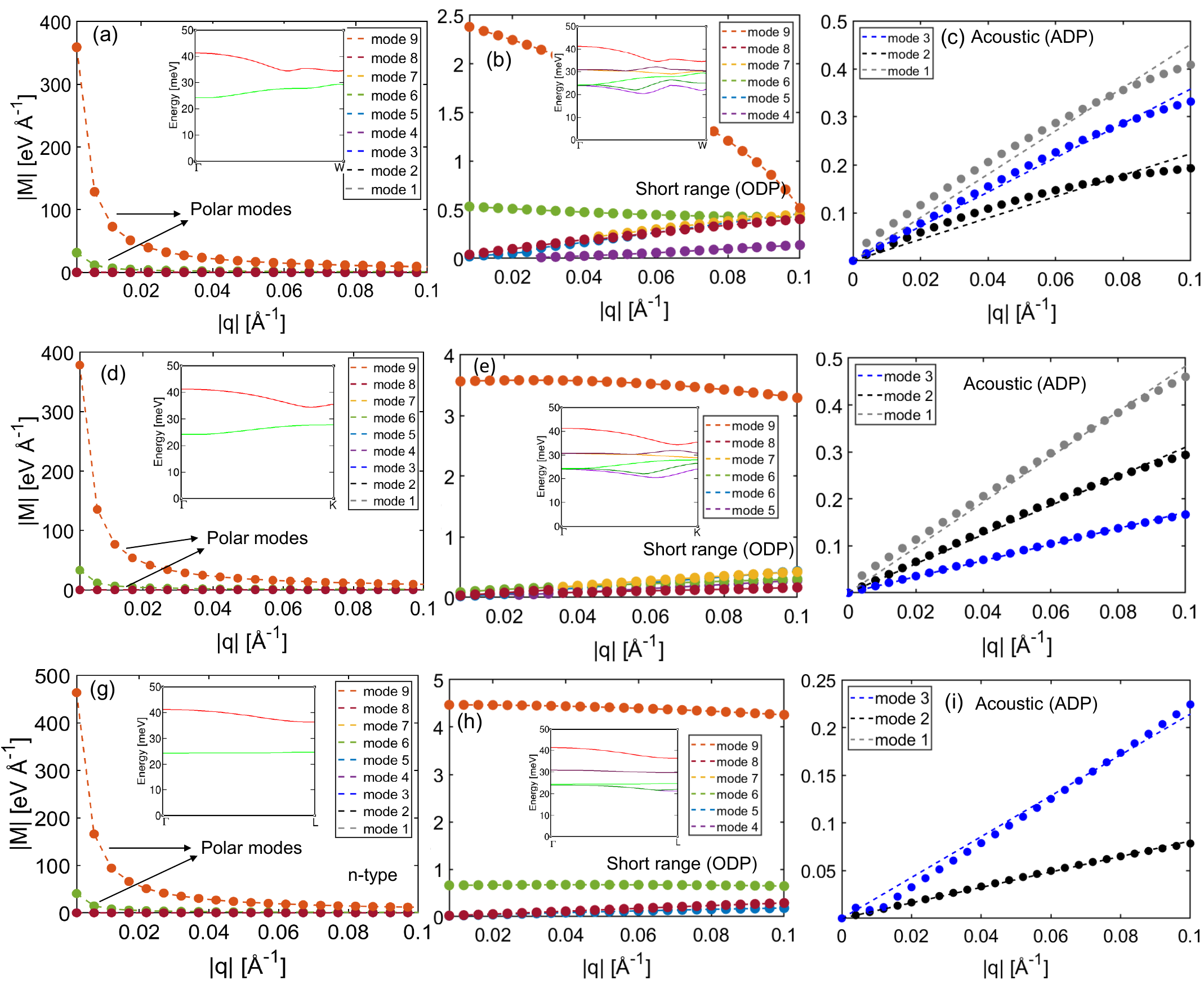}
\caption{Intravalley electron - phonon coupling matrix elements for electrons: The matrix elements in other directions for electrons along (a-c) $\Gamma$- W (d-f) $\Gamma$- K and (g-i) $\Gamma$- K crystallographic directions. (Left (a, d, g)) For all phonons (two modes show polar behaviour) along the $\Gamma$-X line. Middle column (b, e, h): Short range part of matrix elements for optical modes. Right column (c, f, i): acoustic phonon matrix elements for scattering within the valence band maxima (VBM to VBM). The insets in (a, b, d, e, g, h) show the frequencies corresponding to the polar optical and non-polar optical modes respectively.  }
	\label{S4}
\end{figure*}

The intravalley deformation potentials for all four scattering processes (between the two valence bands) along different high-symmetry directions are listed in Table \ref{D_process}. The intravalley scattering matrix elements for transition from band 1$\rightarrow$ band1 along different crystallographic directions are shown in Fig. \ref{S3}.  The intravalley scattering matrix elements for the transition from conduction band minima (CBM) to CBM along different crystallographic directions are shown in Fig. \ref{S4}.

\subsection{Overlap Integrals}
\textcolor{black}{We have calculated the overlap matrix elements using DFT. However, we find that only in the cases where the k-states are fully degenerate the overlap integrals are 1 (intra-band) and 0 (inter-band), as mention by the referee. In fact, for the majority of intra-band and inter-band overlap integrals around the band extrema, the matrix elements are scattered in the entire range from 0 to 1, with an average value of around 0.64. Taking the square of all the values computed and then averaging, we reach an overlap integral squared value to be used for all transitions (intra- and inter-band) of 0.5. We provide details below why these values make sense, which are also included in the revised manuscript and the Supporting information file.}  

\textcolor{black}{For POP scattering, the Fr\"{o}hlich (long‐range) term for \(\mathbf{q}\to 0\)  can be written as: \cite{ivanadefpot}
\begin{equation}
\begin{split}
    g_{mn\nu}^{L}(\mathbf{k},\mathbf{q})
= & 
\frac{ie^{2}}{\varepsilon_{0}\,\Omega}
\braket{ u_{m,\mathbf{k}+\mathbf{q}}}{u_{n,\mathbf{k}}} \times  \\
& \sum_{b}\sqrt{\frac{m_{c}}{m_{b}}}
\sum_{\mathbf{G}\neq -\mathbf{q}}
\frac{(\mathbf{q} + \mathbf{G})\cdot Z_{b}^{*}\cdot \mathbf{e}_{b\nu}(\mathbf{q})}{(\mathbf{q}+\mathbf{G})\cdot\varepsilon_{\infty}\cdot(\mathbf{q}+\mathbf{G})} \times  \\
& e^{-\frac{\lvert \mathbf{q}+\mathbf{G}\rvert^{2}}{4\alpha}}
\end{split}
\end{equation}
where $\braket{ u_{m,\mathbf{k}+\mathbf{q}}}{u_{n,\mathbf{k}}} $ is the overlap of the cell-periodic parts of the Bloch states. Including this overlap factor in \(g_{mn\nu}^{L}\) is a valid approximation to capture the correct weighting of electron–phonon scattering amplitudes as \(\mathbf{q}\to 0\).}

\textcolor{black}{We compute the overlap integrals of the periodic parts as:
\begin{equation}
S_{mn}(\mathbf{k},\mathbf{k}+\mathbf{q})
\;=\;
\bigl\langle u_{m,\mathbf{k}+\mathbf{q}} \mid u_{n,\mathbf{k}}\bigr\rangle,
\end{equation}
directly from the DFT‐calculated wave functions stored in the Quantum ESPRESSO output.
For computational efficiency, and in order to account for the overlap integrals in the entire space around the vicinity of the band extrema, we proceed as follows: We consider the region around the band extrema and place 100 randomised k-points  in the radius of 10\% of the reciprocal lattice vector (approximately 0.1 \AA{}$^{-1}$) from the band extrema.
We then compute the overlap integrals by considering a point as an initial k-state transitioning to all other k-states. We then proceed by considering all points as initial points and all other points as final scattering points.
Thus, we end up computing $\sim N^2$ overlap pairs with maximum distance of 20\% of the Brillouin zone.
We perform this computation for the first two bands of the valence band, for both intra-band and inter-band transitions, and the results are shown, respectively, in Fig. R1.2.1(a,b) for band 1 (upeer valence band - B1) and band 2 (second valence band - B2). The x-axis shows the distance in the 3D k-space that separates the different initial and final states.
We chose to limit our investigation to vectors up to 20\% of the BZ, since both POP and IIS are anisotropic mechanisms and their strength reduces drastically with increasing exchange vectors.
Interestingly, the overlap integral values spread significantly from 0 to 1 with an average of $\sim$0.64 in both the intra-band and inter-band cases for both bands.
Only in the very small k-space distances, the overlaps of wavefunctions tend to be either 1 or 0 for intra- and inter-band transitions (where the bands are almost degenerate); however, this trend disappears quickly as the distance increases.
The scattering rates involve the square of the overlap integral, thus we square these values and afterwards average them, resulting at $\left< S_{mn}(\mathbf{k},\mathbf{k}+\mathbf{q})^2 \right> \approx 0.5$. We then use this single value for all overlap integrals for POP and IIS for the valence band.}

\begin{figure}[htbp!]
  \centering
    \includegraphics[width=0.9\linewidth]{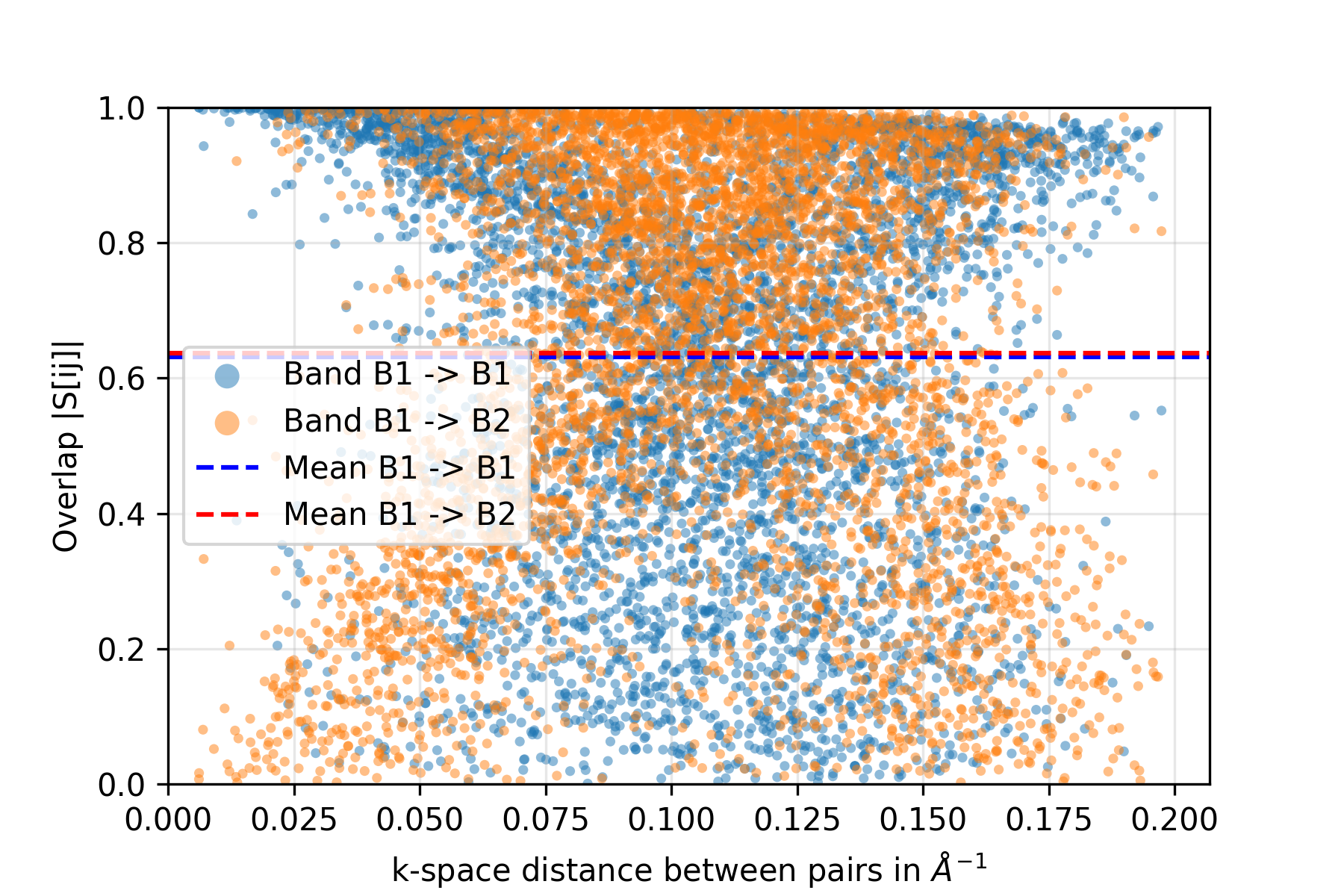}\\
    \includegraphics[width=0.9\linewidth]{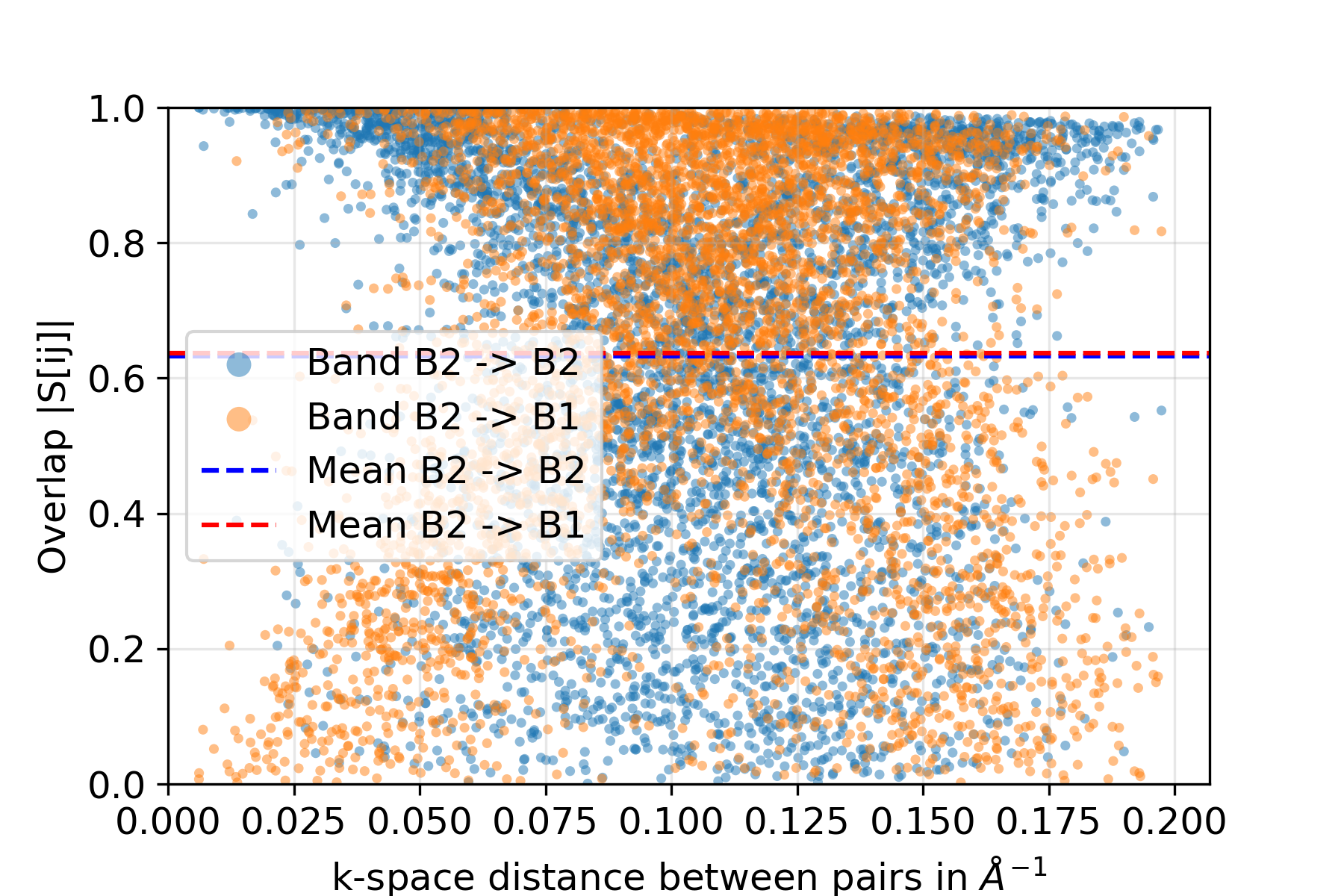}
  \caption{Intra-band (blue) and inter-band (orange) overlap integrals for the first valence band B1 (left) and the second valence band B2 (right). The average values are also indicated. Note that since we place random points in k space uniformly, there are uneven distributions of data sets in pair distances, resulting clustering of points centred around 10 \AA{}$^{-1}$}.
  \label{fig:both_images}
\end{figure}

\begin{figure*}[!htbp]
  \centering
  \includegraphics[width=0.9\textwidth]{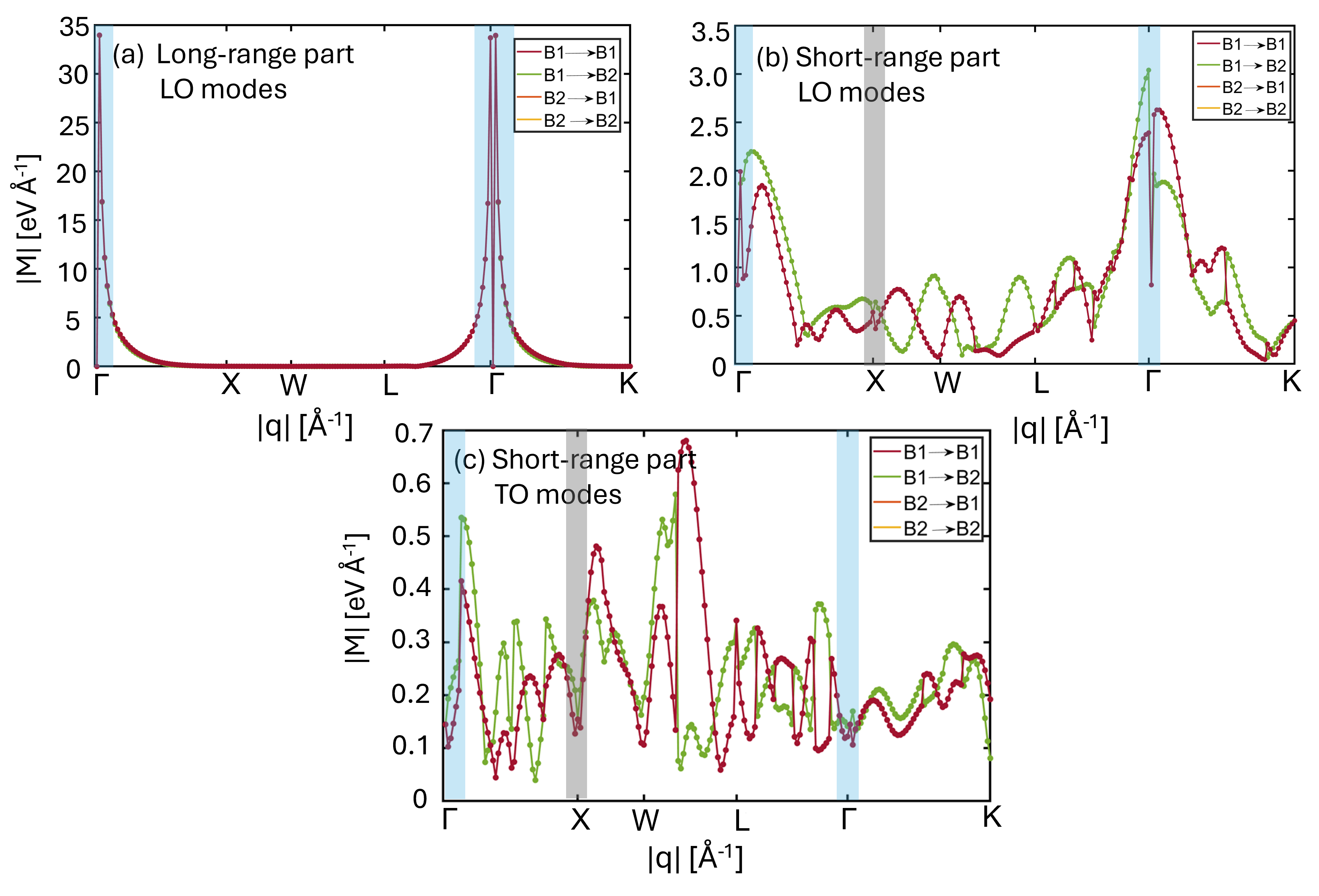} 
  \caption{(a, b) The long-range and short-range parts of matrix elements for longitudinal optical (LO) modes and (c) short range part of matrix elements for transverse optical (TO) modes for valence bands.}
  \label{ML_LO}
\end{figure*}
\begin{figure*}[!htbp]
  \centering
  \includegraphics[width=0.9\textwidth]{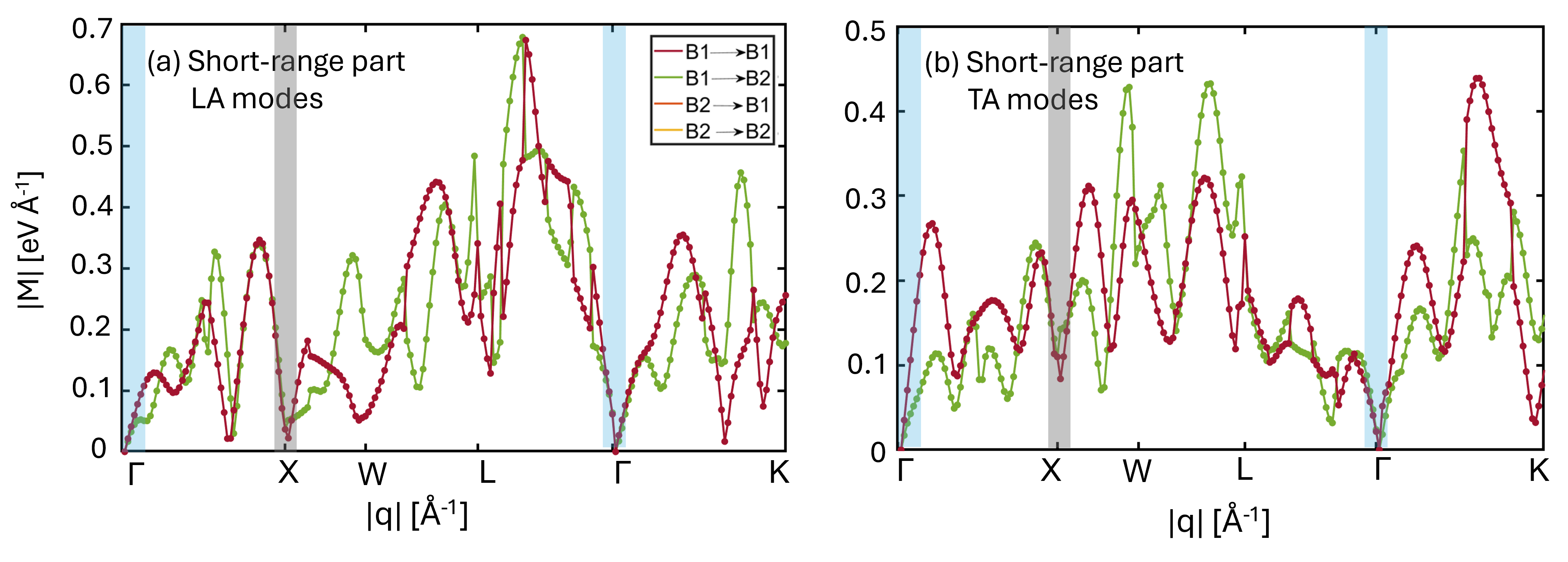} 
  \caption{Short-range part of matrix elements for: (a) longitudinal acoustic (LA), and (b) transverse acoustic (TA) modes for valence bands.}
  \label{ML_short}
\end{figure*}

\begin{figure}[htbp!]
  \centering
    \includegraphics[width=0.9\linewidth]{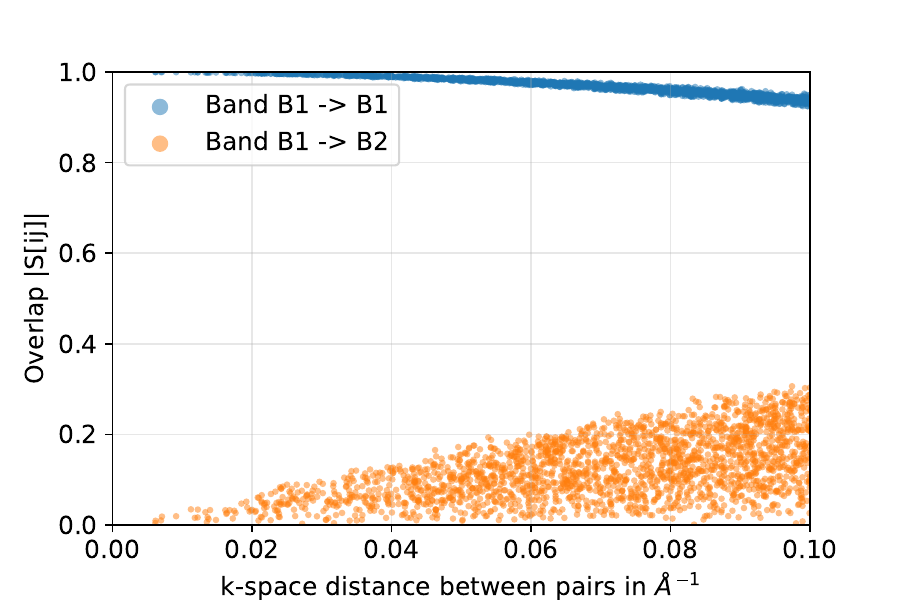}
  \caption{B1-B1 intra-band (blue) and B1-B2 inter-band (orange) overlaps for the conduction band valleys.}
  \label{fig:overlap_X}
\end{figure}

\textcolor{black}{The role of degenerate bands in electron-phonon scattering is partially discussed by Harrison \cite{harrison1956scattering}.
Consider the first-order expansion of the cell periodic part of degenerate Bloch states around the band extremum $\mathbf{k_0}$ using perturbation theory:
\begin{equation}
    \ket{u_{il,\mathbf{k_0+q}}} = \ket{\tilde{u}_{il,\mathbf{k_0}}(\mathbf{k_0+q})} + \sum_{j\neq i;m}\frac{h_{ml}^{ji}(\mathbf{k_0+q})}{E_i-E_j} \ket{u_{jm,\mathbf{k_0}}} 
\end{equation}
where $l$ and $m$ are indices for enumerating degenerate states in the $i$-th and $j$-th eigenstates, respectively. 
Note that, in degenerate perturbation theory, it is necessary to choose the zero-th order term in the expression of the wavefunction $\ket{\tilde{u}_{il,\mathbf{k_0}}}$ as a linear combination of a set of degenerate states of unperturbed wavefunctions $\ket{u_{il,\mathbf{k_0}}} $ at $\mathbf{k_0}$, as:
\begin{equation}
    \ket{\tilde{u}_{il,\mathbf{k_0}}(\mathbf{k_0+q})} = \sum_{l'}c_{il'}(\mathbf{k_0+q})\ket{u_{il',\mathbf{k_0}}}
\end{equation}
such that the perturbation procedures will converge in the first or higher-order terms.
The expression of this `transformed' zero-th order wavefunction is not trivial and depends on each $\mathbf{k_0+q}$.
This implies that, while the degenerate bands on $\mathbf{k_0}$ can be made orthogonal to the perturbed bands on $\mathbf{k_0+q}$, the overlap integral between Bloch states on two points where the degeneracy is lifted, $\mathbf{k_0+q}$ and $\mathbf{k_0+q'}$  does not necessarily vanish in the zero-th order, even if the transition is classified as `inter-band'.
Therefore, based on the degenerate perturbation theory and the presence of overlap integral in the expression of electron-phonon matrix elements, the scattering of electrons between two bands cannot be ruled out (unless it is forbidden by symmetry).} 

\textcolor{black}{To illustrate how the overlap integral changes between degenerate states, let us consider a unitary transformation of two-fold degenerate bands:
\begin{equation}
\begin{pmatrix}
\ket{\tilde{u}_{1,\mathbf{k_0}}} \\
\ket{\tilde{u}_{2,\mathbf{k_0}}}
\end{pmatrix}
=
U_{2\times2}
\begin{pmatrix}
\ket{u_{1,\mathbf{k_0}}} \\
\ket{u_{2,\mathbf{k_0}}}
\end{pmatrix}
\end{equation}
where $U_{2\times2}$ is a 2 by 2 unitary matrix.
If we restrict the expression of $U_{2\times2}$ to be real and its determinant to be $1$, the matrix becomes a proper rotational matrix in a 2 dimensional space:
\begin{equation}
    U_{2\times2} =
    \begin{pmatrix}
        \cos\theta & -\sin\theta \\
        \sin\theta &  \cos\theta \\
    \end{pmatrix}
\end{equation}
which can be parametrised by a single value of $\theta$.
The zeroth-order of the overlap integral between degenerate states and perturbed states is proportional to the corresponding matrix elements of this $U$.
Therefore, the average of the absolute value of overlap integrals for inter- and intra-band transitions is $(1/2\pi)$$\int_0^{2\pi} | \sin \theta | d\theta = (1/2\pi)\int_0^{2\pi} | \cos\theta | d\theta = 2 / \pi  \approx 0.63662$, which also matches the numerical value we computed above.
Similarly, the absolute value of the overlap integral between perturbed states at $\mathbf{k_0 + q}$ and $\mathbf{k_0 + q'}$ can be found.
(However, the average of the square of the overlap integral that we use in the scattering rate expressions would then be $(1/2\pi)$$\int_0^{2\pi}  \sin^2 \theta  d\theta = 0.5$)}

\textcolor{black}{It is also worth mentioning that, in practical DFT calculations, coefficients of basis sets among degenerate bands are somewhat rather arbitrary, as any unitary transformation of degenerate wavefunctions leave the eigenvalues of Kohn-Sham Hamiltonian invariant.
However, this unitary transformation of Bloch states will change their electron-phonon matrix elements, and different DFT calculations may yield different values.
While there are approaches to 'fix' the arbitrariness of the rotation by choosing the coefficients of specific basis to be real and making it transform with symmetry operations to circumvent this issue, one may also choose to calculate the average of squares of the matrix elements, over the degenerate electronic bands (as well as degenerate phonon branches) in post-processing analysis.
These values are shown in Fig. \ref{fig:both_images} for intra- and inter-band transitions in the valence band (i.e. transitions from B1-B1 and B1-B2) as distinct points.}

\textcolor{black}{This is also the behaviour of QuantumESPRESSO and EPW, when a user requests an output of each individual matrix element from DFPT or Wannier interpolations.
In this way, there will be no distinction between two bands, and both 'intra-band' and 'inter-band' matrix elements for a given set of degenerate bands will be identical to each other.
The long range part of matrix elements for LO modes extracted from EPW are the same between B1-B1 and B1-B2, or B2-B1 and B2-B2 as shown in Fig. \ref{ML_LO}.
The short range part of the matrix elements for transitions from B1-B1 and B2-B1 (with same final state) for LO and TO modes are the same, and similarly matrix elements for transitions from B1-B2 and B2-B2 are the same as shown in Fig. \ref{ML_short}.
For this reason, we have also opted to take a similar approach by using the same matrix elements for both intra- and inter-band scattering process, and weight the scattering rate by the average of the squares of the overlap integrals in both processes.}

\textcolor{black}{We have also performed this analysis for the conduction band.
In this case only one band appears at the band extrema at the X-point, whereas the second band is much higher in energy.
The intra-band overlap integrals are close to unity, whereas the inter-band overlap integrals are more or less negligible (Fig. \ref{fig:overlap_X}).
This information is also included in the simulations.
Finally, note that no extra considerations are needed for the overlap integrals in the case of the ADP and ODP scattering processes, since this information is already included in the matrix elements (see Fig. \ref{ML_short}) for the matrix elements for valence bands, which follow the behaviour described above).}

\subsection{Input Parameters for BTE calculation}
The required BTE input parameters i.e. density $\rho$, sound velocity u$\textsubscript{s}$, the velocities of acoustic phonon modes ( v$\textsubscript{LA}$ (for longitudinal), v$\textsubscript{TA}$ and   v$\textsubscript{TA’}$ (for two transverse)) and the static and high frequency dielectric constants obtained through first-principles calculation are listed in Table \ref{parameters}.

\begin{table*}[!htbp]
\centering
 \begin{tabular}{|c|c|c|c|c|c| }
    \hline
$\rho$ (kg/$m^3\times10^3$)	&u$\textsubscript{s}$(m/s$\times10^3$)	&v$\textsubscript{LA}$(m/s$\times10^3$)	&v$\textsubscript{TA}$(m/s$\times 10^3$)	&v$\textsubscript{TA’}$(m/s$\times10^3$)	&$k_s/k_\infty$ \\
\hline
8.45	&3.4	&5.1	&3.1	&2.9	&42.83/23.91  \\
 \hline
\end{tabular}
 \caption{The required BTE input parameters dielectric constants and mass densities, which are also obtained through first-principles calculations.}
  \label{parameters}
\end{table*}
\begin{figure*}[!htbp]
	\centering
	\includegraphics[width=0.8\textwidth]{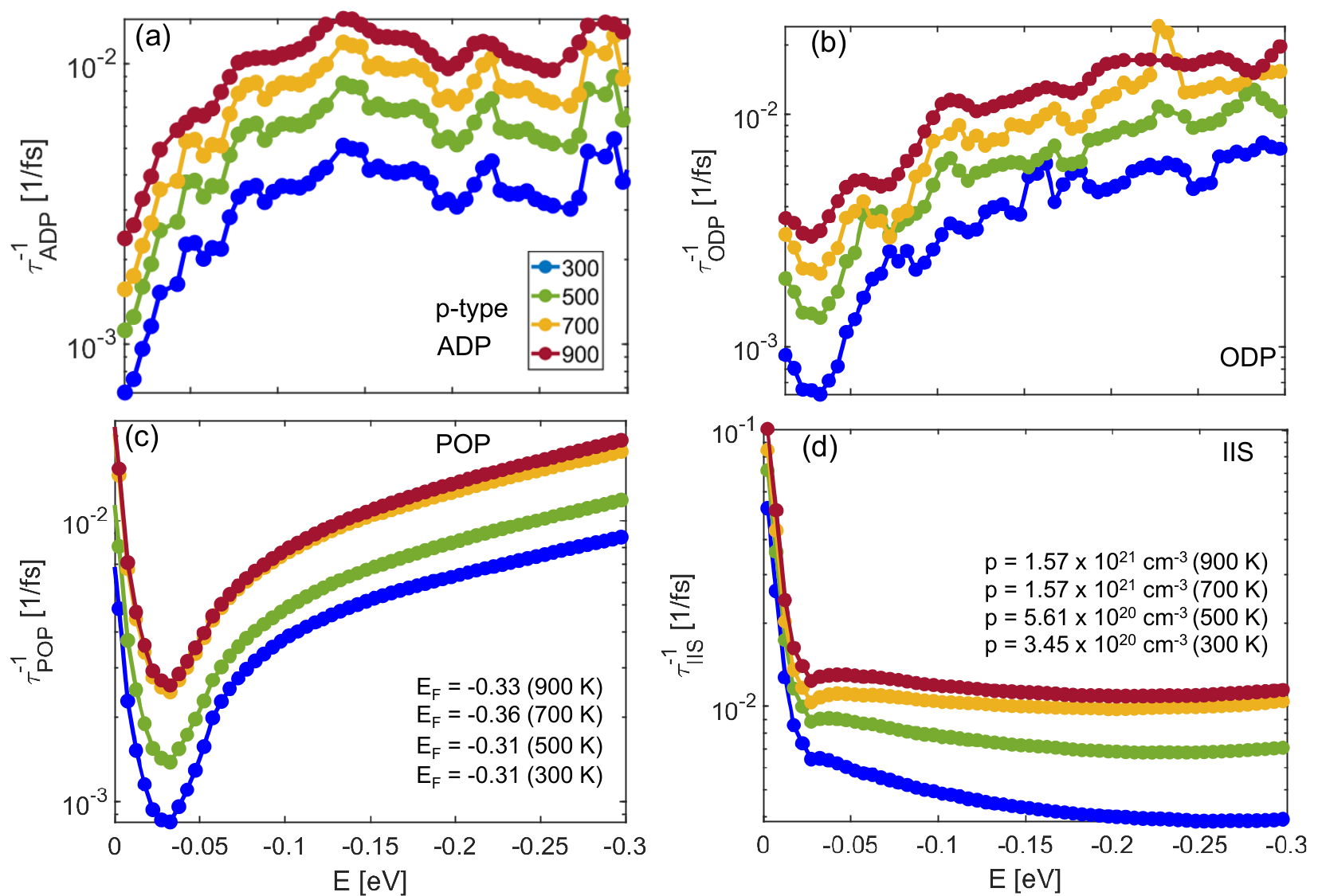}
\caption{The hole scattering rates for different scattering processes versus energy and for different temperatures as noted. (a) ADP. (b) ODP. (c) POP. (d) IIS}
	\label{S5}
\end{figure*}
\newpage
\subsection{Electron and hole Scattering rates at different temperatures}
\begin{figure*}[!htbp]
	\centering
	\includegraphics[width=0.8\textwidth]{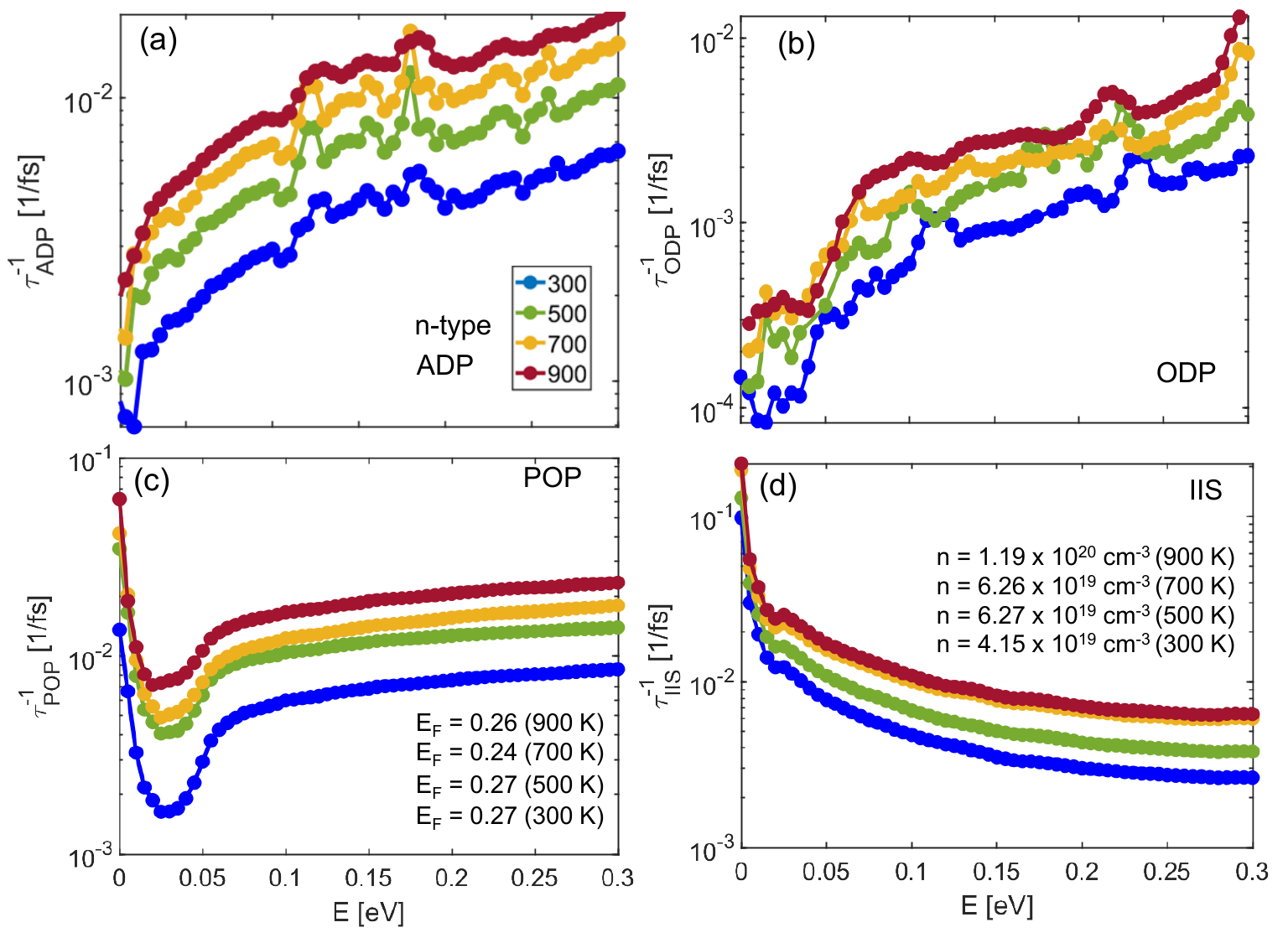}
\caption{The electron scattering rates for different scattering processes versus energy and for different temperatures as noted. (a) ADP. (b) ODP. (c) POP. (d) IIS}
	\label{S6}
\end{figure*}

The hole and electron scattering rates for different processes (ADP, ODP, POP and IIS) as a function of energy and temperature are shown in Fig. \ref{S5} and \ref{S6} respectively. The POP and IIS scattering rates are plotted at the Fermi level corresponding to the maximum value of the power factor at that temperature. 
\textcolor{black}{The scattering rates for each mechanism increase with temperature. }

\begin{figure*}[!htbp]
	\centering
	\includegraphics[width=0.7\textwidth]{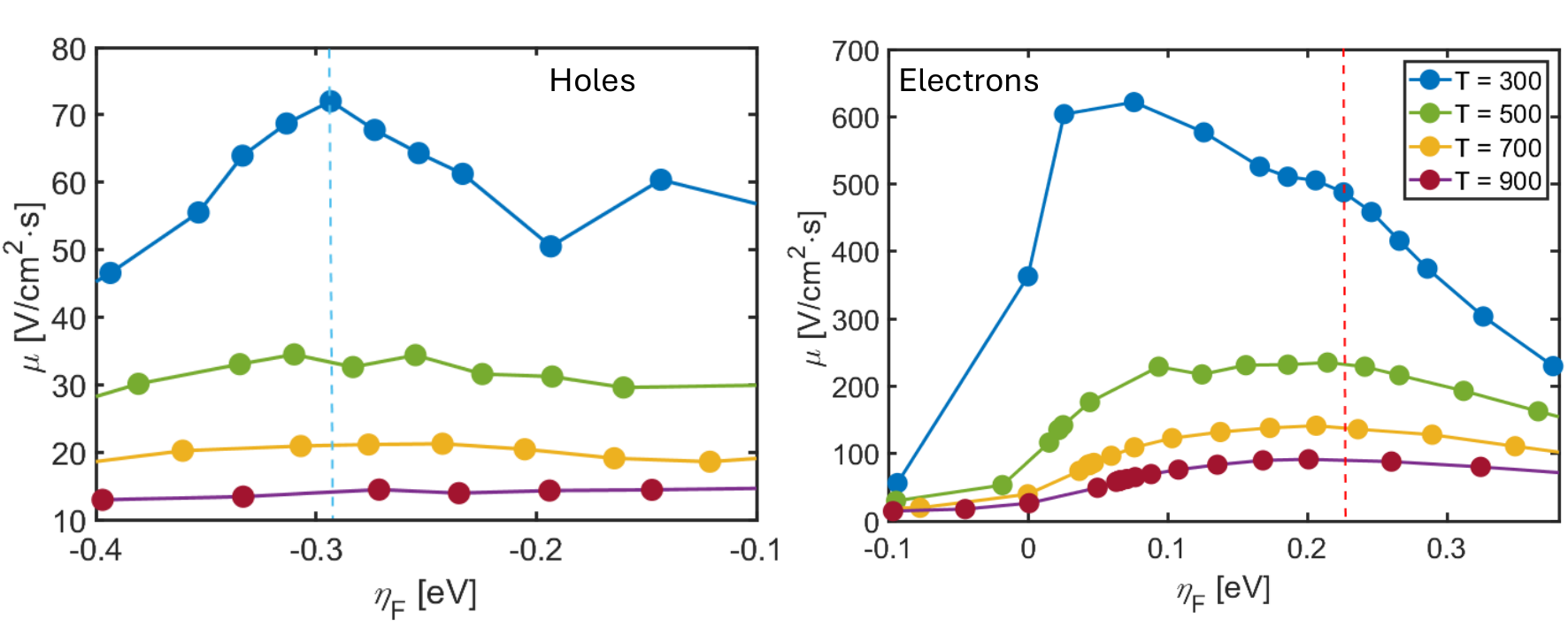}
\caption{The hole and electron mobility vs the fermi level considering all the relevant scattering mechanisms (ADP, ODP, POP, IIS). Results for various temperatures are shown. The blue and red lines are valence and conduction band edges respectively.}
	\label{S7}
\end{figure*}

\newpage
\subsection{Mobility}
The mobility of holes and electrons as a function of temperature is shown in Fig. \ref{S7}. The bipolar effects are included, and the hole and electron mobilities are split into two figures because of the difference in y-scales. The electron mobility values are higher than those of the holes, which is expected due to the large difference in the effective mass of holes and electrons in NbFeSb.

\newpage
\renewcommand\refname{References}
\bibliography{rsc} 
\bibliographystyle{rsc}

\end{document}